\documentclass[preprint,12pt,sort&compress]{elsarticle}

\usepackage{amssymb}
\usepackage{amsmath}

\begin{document}
	
	\begin{frontmatter}
		
		\title{A Voxel-Based Quantum Computing Method (VBQC) for Solid Mechanics Problem} 
		
		\author[1]{Feng Wu} 
		\author[1]{Yuxiang Yang}
		\author[1]{Li Zhu}
		\author[1]{Chen Li}
		\author[1]{Yansong Guo}
		\author[1]{Xu Guo\corref {cor1}}
		\ead{guoxu@dlut.edu.cn}
		
		\cortext[cor1]{Corresponding author} 
		
		\affiliation[1]{organization={State Key Laboratory of Structural Analysis, Optimization and CAE Software for Industrial Equipment, School of Mechanics and Aerospace Engineering},
			addressline={Dalian University of Technology}, 
			city={Dalian},
			postcode={116024}, 
			state={Liaoning},
			country={P. R. China}}
		
		\begin{abstract}
			Quantum computing presents a promising method to overcome the efficiency and memory constraints in large-scale mechanical problems, with numerous successful applications demonstrated in fluid mechanics. However, solid mechanics problems usually require irregular grids for spatial discretization, due to the Lagrange formulations and complex boundaries, which makes the quantum simulation of the system matrix, e.g., the mass or stiffness matrix which is often referred to as the Hamiltonian in quantum computing, difficult to be effectively conducted. This study proposes a voxel-based quantum computing method (VBQC) for the quantum simulation of Hamiltonians in solid mechanics. VBQC applies voxel grids to discretize the spatial domain, thereby enabling the system matrix to exhibit the tridiagonal fractal property. Based on this property, the system matrix can be decomposed into three groups of fundamental matrices, $\mathbf{k}_{n}$, $\mathbf{c}_{n}$, and $\mathbf{q}_{n}$. This decomposition process is referred to as the KCQ decomposition. By integrating the KCQ decomposition with the quantum Fourier transform and the quantum multiplexer, VBQC enables efficient quantum simulation of Hamiltonians in solid mechanics. Three specific solid problems with different dimensions and numbers of variables are applied to preliminarily verify the correctness of the proposed VBQC for solid mechanics problems.
		\end{abstract}

		
	\end{frontmatter}
	
		\section{Introduction}
		\label{S1}
		Quantum computing is currently developing rapidly and has been successfully applied in fields such as computational chemistry, fluid mechanics, machine learning, and so on \citep{R4,R5,RB1,R1,RB2}. In computational mechanics, an increasing number of studies on quantum algorithms have emerged, with the following categories of work being particularly noteworthy: research on the quantum representation theory of mechanical problems to accommodate quantum computing rules, typically involving studies on how to Schrödingerize mechanical problems and then solve them using Hamiltonian quantum simulation within the Schrödinger framework \citep{R2,R3,R4,R5,R6,R7,RB3}; using basic quantum algorithms, such as quantum phase estimation, quantum amplitude estimation, the Harrow-Hassidim-Lloyd (HHL) algorithm, and so on, to solve problems widely involved in computational mechanics, such as matrix eigenvalues and systems of linear equations, to achieve quantum acceleration \citep{R8,R9,R10}; research on quantum-classical hybrid computing methods, which generally transform the solution of systems of equations in computational mechanics into optimization problems and then use quantum optimization methods such as variational quantum algorithms and quantum annealing for solution \citep{R11,R12,R13,R14,R15,RB4}; employing quantum Fourier transform method to solve computational mechanics problems \citep{R11,R16}, which are currently mainly applicable to mechanical problems in regular solution domains, with relatively little research on handling irregular solution domains or spatially varying physical parameters within the domain; and leveraging the exponential advantage of quantum algorithms in distance calculations to innovatively propose quantum-enhanced data-driven computational mechanics methods \citep{RBHu1,RBHu2}, among others. Except for the quantum annealing algorithm, which has been successfully applied in the field of topology optimization and is based on quantum annealers \citep{R15}, all other algorithms rely on quantum gate-based quantum computers. Fundamental quantum algorithms such as quantum simulation of Hamiltonians, quantum phase estimation, the HHL algorithm, and variational quantum algorithms all involve a core step: matrix decomposition \citep{R12,R17,R18}. However, due to the adoption of Lagrangian descriptions and the prevalence of complex boundary conditions in solid mechanics problems, how to effectively decompose the system matrix (e.g., stiffness matrix) remains one of the pressing challenges in quantum computation for solid mechanics.
		
		After spatially discretizing a mechanical problem using finite elements or other algorithms such as meshfree methods, the resulting system matrix, denoted as $\mathbf{A}\in {{\mathbb{C}}^{{{2}^{n}}\times {{2}^{n}}}}$, is generally a Hermitian matrix \citep{R19}. In the quantum simulation of the Schrödinger equation, quantum phase estimation, and the HHL algorithm, a critical step is efficiently computing ${{e}^{\text{i}\mathbf{A}t}}$ using a quantum simulator, which is commonly referred to as the quantum simulation of the Hamiltonian (QSH). The efficient implementation of QSH \citep{R20} relies on decomposing the matrix $\mathbf{A}$ into a series of submatrices ${{\mathbf{A}}_{k}}$ that can be quantum-efficiently simulated. Current research on decomposing Hermitian matrix $\mathbf{A}$ is mainly based on two categories: One is decomposing $\mathbf{A}$ into a linear combination of unitary matrices (LCU) \citep{R21}, expressed as $\mathbf{A}=\sum\limits_{i=1}^{Q}{{{a}_{i}}{{\mathbf{U}}_{i}}}$. LCU decomposition is not only a key step in the QSH but also critical for implementing other quantum algorithms such as variational quantum algorithms \citep{R13}. LCU decomposition typically requires two prerequisites: 1) $\mathbf{A}$ can be decomposed into $Q=\text{poly}\left( n \right)$ unitary matrices ${{\mathbf{U}}_{i}}$; 2) ${{e}^{\text{i}{{\mathbf{U}}_{i}}t}}$ can be efficiently simulated, i.e., achievable with $\text{poly}\left( n \right)$ quantum gate operations. However, for general Hermitian matrices $\mathbf{A}$, there is currently no universal method to satisfy these two conditions. A relatively mature decomposition is the Pauli decomposition, but the number of terms $Q$ in this method can be as high as ${{4}^{n}}$, eliminating the advantages of quantum computing. The other category is decomposing $\mathbf{A}$ into a series of 1-sparse Hermitian matrices ${{\mathbf{A}}_{k}}$, where each row and column of ${{\mathbf{A}}_{k}}$ contains only one non-zero element \citep{R22}. It can be proven that an s-sparse Hermitian matrix $\mathbf{A}$ (with at most s non-zero elements per row/column) can theoretically be decomposed into s+1 1-sparse Hermitian matrices. However, whether the matrix exponential ${{e}^{\text{i}{{\mathbf{A}}_{k}}t}}$ of a 1-sparse Hermitian matrix can be efficiently simulated remains a conjecture, and how to construct a quantum algorithm for this purpose remains an open problem.
		
		It can be stated that further in-depth research is still needed on how to utilize the mathematical and physical properties of the system matrix $\mathbf{A}$ obtained through spatial discretization to construct efficient decomposition methods for effective quantum simulation. Current studies on quantum computing in solid mechanics either explore the application of quantum algorithms in mechanical problems based on the premise that $\mathbf{A}$ can be effectively decomposed and simulated, or leverage the characteristics of specific mechanical problems to construct simulations of ${{e}^{\text{i}\mathbf{A}t}}$ for quantum emulation. In summary, general methods for the QSH (i.e., ${{e}^{\text{i}\mathbf{A}t}}$) in solid mechanics problems still remain underdeveloped in quantum computing research. Voxel grids, one of the most commonly used grids in the early development of computational mechanics, offer advantages such as simple modeling, uniform element matrices, and ease of parallelization, making them a promising breakthrough for achieving quantum simulation of the Hamiltonian in general solid mechanics problems. The concept of voxel grids traces back to the finite difference method, which typically employs regular grids and constructs corresponding numerical discretization schemes \citep{R23,R24}. In the finite element method, regular voxel grids (e.g., quadrilateral or hexahedral grids) are also widely used for spatial discretization in solving solid mechanics problems. However, voxel grids cannot perfectly conform to curves or surfaces, leading to poor accuracy in handling complex boundaries. To address the issue, methods such as the octree subdivision method \citep{RBFCM1} have been developed. Based on this, \citet{RBFCM2,RBFCM1} introduced the fictitious domain concept and proposed the finite cell method, which has been successfully applied in simulating heterogeneous materials \citep{RBVBY} and bone structures \citep{RBVBG}. Additionally, the implicit boundary method \citep{RBIBM1,RBIBM2}, the fat boundary method \citep{RBFBM}, and other methods \citep{RBWBM} have been proposed to address the issue that voxel grids are difficult to precisely simulate complex boundaries. Recently, Guo et al. \citep{R25,R26,R27} combined machine learning with computational mechanics and proposed the problem-independent machine learning method that achieves two orders of magnitude higher computational efficiency than traditional finite elements, enabling efficient analysis of finite elements under ultra-large-scale voxel grids. It indicates that with appropriately designed computational frameworks, efficient and high-precision finite element simulations using large-scale voxel grids are achievable, significantly reducing preprocessing time. In addition, quantum computing utilizes quantum superposition and entanglement properties and therefore naturally has exponential parallelism and storage capacity, with excellent compatibility with voxel grids. Take a ${{2}^{n}}\text{-dimensional}$ vector $\mathbf{u}=\left( {{u}_{0}},\ {{u}_{1}},\ \cdots ,\ {{u}_{{{2}^{n}}-1}} \right)$ as an example: the classical computer requires $O\left( {{2}^{n}} \right)$ bits to store the information and at least $O\left( {{2}^{n}} \right)$ operations to compute on the vector, while the quantum computer can encode the vector using $n$ qubits to obtain the quantum state $\left| u \right\rangle =\sum\limits_{i=0}^{{{2}^{n}}-1}{{{u}_{i}}\left| i \right\rangle }$, often requiring only $\text{poly}\left( n \right)$ operations for computations \citep{R22}. Therefore, voxel grids are naturally suitable for quantum computing: on the one hand, using voxel grids to model mechanical problems on quantum computers eliminates storage limitations; on the other hand, voxel grids offer greater promise for constructing effective quantum algorithms for general solid mechanics problems.
			
		This study attempts to investigate quantum simulation in mechanical problems, primarily aiming to propose a method for effectively decomposing the Hermitian matrix $\mathbf{A}$ in general solid mechanics problems (e.g., those involving inhomogeneous materials and irregular boundary conditions), and further propose the VBQC for QSH (i.e., ${{e}^{\text{i}\mathbf{A}t}}$). Based on the above study, VBQC is employed in combination with classical quantum algorithms to conduct solution analysis for general solid mechanics problems. The main contributions of this paper are as follows: 1) The fractal property of system matrices generated by voxel grids (MVG) in mechanical problem modeling are investigated. Based on the property, the KCQ decomposition for MVG is proposed by comprehensively utilizing circulant matrices, matrix direct products, direct sums, Pauli matrices, etc. The KCQ decomposition method decomposes the MVG into three groups of fundamental matrices ${{\mathbf{k}}_{n}}$, ${{\mathbf{c}}_{n}}$, and ${{\mathbf{q}}_{n}}$ with dimensions of $d=1,\ 2,\ \text{or}\ \text{4}$, thereby decomposing ${{e}^{\text{i}\mathbf{A}t}}$ into the computation of the exponents of these three groups of $d\text{-dimensional}$ matrices. 2) Based on the KCQ decomposition and using quantum Fourier transform (QFT), circulant gates, phase gates, and quantum multiplexers, the VBQC for the exponents of MVG is proposed, and corresponding quantum circuits are designed. The basic introduction to quantum computing is provided in section \ref{S2}. Section \ref{S3} proposes using voxel grids to discretize mechanical problems, studies the tridiagonal fractal property of the MVG, and develops the KCQ decomposition for the MVG by using the tridiagonal fractal property. Section \ref{S4} proposes the VBQC for QSH based on the KCQ decomposition, with the constructed quantum circuits applicable to solid mechanics problems of arbitrary dimensions. Before the final section concludes the paper, section \ref{S5} presents three numerical examples to verify the correctness of the proposed method.
			
		\section{Basic Introduction to Quantum Computing}
		\label{S2}		
		This section briefly introduces relevant quantum computing knowledge used in the subsequent content \citep{R22,R28}.
		
		\subsection{Basic Quantum Gates}
		\label{S2S1}
		Classical computers operate with bits, which have two states: 0 or 1, while quantum computers operate through operations on qubits. The state of a qubit, which can be 0, 1, or a superposition of both, is generally denoted as $\left| q \right\rangle ={{a}_{0}}\left| 0 \right\rangle +{{a}_{1}}\left| 1 \right\rangle $ where ${{a}_{0}}$ and ${{a}_{1}}$ are complex numbers satisfying ${{\left| {{a}_{0}} \right|}^{2}}+{{\left| {{a}_{1}} \right|}^{2}}=1$. Here, ${{\left| {{a}_{0}} \right|}^{2}}$ and ${{\left| {{a}_{1}} \right|}^{2}}$ represent the probabilities of observing the quantum state $\left| q \right\rangle $ as $\left| 0 \right\rangle $ or $\left| 1 \right\rangle $, respectively. A quantum system can consist of multiple qubits. For example, a system with two qubits can be in a state expressed as $\left| q \right\rangle ={{a}_{0}}\left| 00 \right\rangle +{{a}_{1}}\left| 01 \right\rangle +{{a}_{2}}\left| 10 \right\rangle +{{a}_{3}}\left| 11 \right\rangle $, which can exist in four basis states: $\left| 00 \right\rangle $, $\left| 01 \right\rangle $, $\left| 10 \right\rangle $, and $\left| 11 \right\rangle $. The coefficients measure the probabilities of observing the basis states when the quantum state $\left| q \right\rangle $ is measured. If a 2-qubit state $\left| q \right\rangle $ can be expressed as the direct product of two single-qubit states $\left| x \right\rangle $ and $\left| y \right\rangle $:
		\begin{equation}
			\left| q \right\rangle =\left| x \right\rangle \otimes \left| y \right\rangle.
			\label{E1}
		\end{equation}
		then $\left| q \right\rangle $ is called a separable state; otherwise, it is an entangled state.
		
		In quantum computing, computational processes are primarily implemented through unitary transformations on quantum states. A classical example of the unitary transformation is the Pauli-X gate whose matrix representation is
		\begin{equation}
			\mathbf{X}=\left[ \begin{matrix}
				0 & 1  \\
				1 & 0  \\
			\end{matrix} \right].
			\label{E2}
		\end{equation}
		The unitary transformation acts on a single qubit $\left| q \right\rangle ={{a}_{0}}\left| 0 \right\rangle +{{a}_{1}}\left| 1 \right\rangle $, yielding:
		\begin{equation}
			\mathbf{X}\left| q \right\rangle ={{a}_{0}}\mathbf{X}\left| 0 \right\rangle +{{a}_{1}}\mathbf{X}\left| 1 \right\rangle ={{a}_{0}}\left| 1 \right\rangle +{{a}_{1}}\left| 0 \right\rangle .
			\label{E3}
		\end{equation}
		The action of the Pauli-X gate is equivalent to flipping the qubit state $\left| 0 \right\rangle $ to $\left| 1 \right\rangle $ and $\left| 1 \right\rangle $ to $\left| 0 \right\rangle $. Several common single-qubit gates are listed in Table \ref{T1}.
		\begin{table}[!h]
			\caption{Common Quantum Gates and The Corresponding Circuits}
			\label{T1}
			\renewcommand\arraystretch{1} 
			\setlength{\tabcolsep}{5mm}
			\centering
			\begin{tabular}{ccc}
				\hline
				Quantum gate & Matrix form & Quantum circuit \\ 
				\hline 
				$\begin{matrix}
					{}\\
					\text{Hadamard gate}\\
					{}
				\end{matrix}$
				&
				$\begin{matrix}
					{}\\
					\mathbf{H}=\frac{1}{\sqrt{2}}\begin{bmatrix}
						1 & 1  \\
						1 & -1  \\
					\end{bmatrix}\\
					{}
				\end{matrix}$
				&
				$\begin{matrix}
					{}\\
					\begin{minipage}[b]{0.3\columnwidth}
						\centering
						\raisebox{-.5\height}{\includegraphics[width=\linewidth]{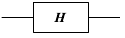}}
					\end{minipage}\\
					{}
				\end{matrix}$ \\
				$\begin{matrix}
					{}\\
					\text{Pauli-X gate}\\
					{}
				\end{matrix}$
				&
				$\begin{matrix}
					{}\\
					\mathbf{X}=\begin{bmatrix}
						0 & 1  \\
						1 & 0  \\
					\end{bmatrix}\\
					{}
				\end{matrix}$
				&
				$\begin{matrix}
					{}\\
					\begin{minipage}[b]{0.3\columnwidth}
						\centering
						\raisebox{-.5\height}{\includegraphics[width=\linewidth]{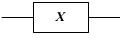}}
					\end{minipage}\\
					{}
				\end{matrix}$\\
				$\begin{matrix}
					{}\\
					\text{Pauli-Y gate}\\
					{}
				\end{matrix}$
				&
				$\begin{matrix}
					{}\\
					\mathbf{Y}=\begin{bmatrix}
						0 & -\text{i}  \\
						\text{i} & 0  \\
					\end{bmatrix}\\
					{}
				\end{matrix}$
				&
				$\begin{matrix}
					{}\\
					\begin{minipage}[b]{0.3\columnwidth}
						\centering
						\raisebox{-.5\height}{\includegraphics[width=\linewidth]{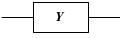}}
					\end{minipage}\\
					{}
				\end{matrix}$\\
				$\begin{matrix}
					{}\\
					\text{Pauli-Z gate}\\
					{}
				\end{matrix}$
				&
				$\begin{matrix}
					{}\\
					\mathbf{Z}=\begin{bmatrix}
						1 & 0  \\
						0 & -1  \\
					\end{bmatrix}\\
					{}
				\end{matrix}$
				&
				$\begin{matrix}
					{}\\
					\begin{minipage}[b]{0.3\columnwidth}
						\centering
						\raisebox{-.5\height}{\includegraphics[width=\linewidth]{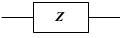}}
					\end{minipage}\\
					{}
				\end{matrix}$\\
				$\begin{matrix}
					{}\\
					\text{Phase gate}\\
					{}
				\end{matrix}$
				&
				$\begin{matrix}
					{}\\		
					\mathbf{T}\left( \theta  \right)=\begin{bmatrix}
						1 & 0  \\
						0 & {{e}^{\text{i}\theta }}  \\
					\end{bmatrix}\\
					{}
				\end{matrix}$
				&
				$\begin{matrix}
					{}\\
					\begin{minipage}[b]{0.3\columnwidth}
						\centering
						\raisebox{-.5\height}{\includegraphics[width=\linewidth]{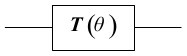}}
					\end{minipage}\\
					{}
				\end{matrix}$\\
				$\begin{matrix}
					{}\\
					\text{Rotation gate}\\
					{}
				\end{matrix}$
				&
				$\begin{matrix}
					{}\\					
					\mathbf{R}\left( \theta  \right)=\begin{bmatrix}
						\cos \theta  & -\sin \theta   \\
						\sin \theta  & \cos \theta   \\
					\end{bmatrix}\\
					{}
				\end{matrix}$
				&
				$\begin{matrix}
					{}\\
					\begin{minipage}[b]{0.3\columnwidth}
						\centering
						\raisebox{-.5\height}{\includegraphics[width=\linewidth]{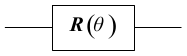}}
					\end{minipage}\\
					{}
				\end{matrix}$\\
				\hline
			\end{tabular}
		\end{table}
		
\begin{figure}[h]
			\centering
			\includegraphics{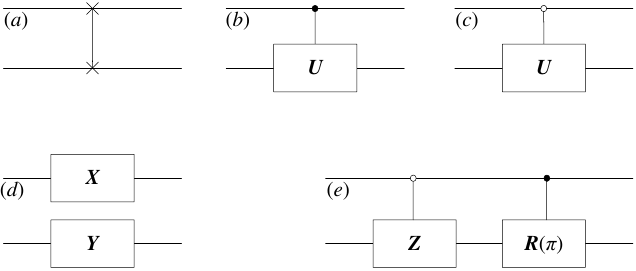}
			\caption{Quantum circuits of the commonly used two-qubit gates: (a) swap gate; (b) controlled-U gate; (c) zero-controlled-U gate; (d) $\mathbf{X}\otimes \mathbf{Y}$ gate; (e) two-qubit multiplexer.}
			\label{F1}
		\end{figure}

		The aforementioned quantum gates are single-qubit gates. In addition, there are several commonly used two-qubit gates, which will be introduced. The first one is the swap gate:
		\begin{equation}
			{{\mathbf{U}}_{\text{swap}}}=\left[ \begin{matrix}
				1 & 0 & 0 & 0  \\
				0 & 0 & 1 & 0  \\
				0 & 1 & 0 & 0  \\
				0 & 0 & 0 & 1  \\
			\end{matrix} \right] .
			\label{E4}
		\end{equation}
		When the swap gate acts on a two-qubit system, it swaps the states of the two qubits: $\left| 00 \right\rangle \to \left| 00 \right\rangle $,$\left| 01 \right\rangle \to \left| 10 \right\rangle $, $\left| 10 \right\rangle \to \left| 01 \right\rangle $, $\left| 11 \right\rangle \to \left| 11 \right\rangle $. Another commonly used two-qubit gate is the controlled-U gate, which is defined as ${{\mathbf{I}}_{2}}\oplus \mathbf{U}$, where $\oplus $ denotes matrix direct sum. The controlled-U gate functions as a conditional operation: If the first qubit, which is called the control qubit, is in the state $\left| 0 \right\rangle $, the state of the second qubit, which is called the target qubit, remains unchanged; and if the first qubit is in the state $\left| 1 \right\rangle $, the unitary gate $\mathbf{U}$ is applied to the second qubit. The condition can also be exchanged: if the first qubit is $\left| 1 \right\rangle $, the second qubit remains unchanged; if the first qubit is $\left| 0 \right\rangle $, the unitary gate $\mathbf{U}$ is applied to the second qubit. This variant is often called the zero-controlled-U gate. The quantum circuits of the swap gate, controlled-U gate, and zero-controlled-U gate are shown in Fig. \ref{F1}.
		
		\subsection{Quantum Parallelism}
		\label{S2S2}
		In quantum computing, quantum parallelism refers to two main scenarios. The first involves the inherent parallelism of basic quantum gates. Taking a rotation gate as an example, applying a single rotation gate to a general quantum state $\left| q \right\rangle ={{a}_{0}}\left| 0 \right\rangle +{{a}_{1}}\left| 1 \right\rangle$ transforms it into:
		\begin{equation}
			\mathbf{R}\left| q \right\rangle =\left( \cos \theta {{a}_{0}}-\sin \theta {{a}_{1}} \right)\left| 0 \right\rangle +\left( \sin \theta {{a}_{0}}+\cos \theta {{a}_{1}} \right)\left| 1 \right\rangle ,
			\label{E5}
		\end{equation}
		which means a single rotation gate operation effectively computes both $\cos \theta {{a}_{0}}-\sin \theta {{a}_{1}}$ and $\sin \theta {{a}_{0}}+\cos \theta {{a}_{1}}$ simultaneously. The second scenario involves the direct product of quantum gates. For a two-qubit system, quantum gates are typically represented by $4\times 4$ matrices. If a unitary matrix is the direct product of two basic quantum gates, quantum parallelism enables their parallel operation. For example, consider a $4\times 4$ quantum gate $\mathbf{U}=\mathbf{X}\otimes \mathbf{Y}$, whose circuit is shown in Fig. \ref{F1} (d). Although this circuit contains two basic gates, each gate acts independently on a separate qubit without interacting with each other, allowing the two gates to operate in parallel.
		
		\subsection{Quantum Multiplexer}
		\label{S2S3}
\begin{figure}[h]
			\centering
			\includegraphics{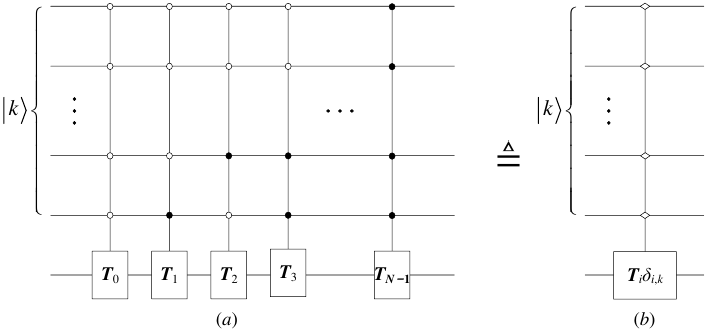}
			\caption{Quantum circuits of the quantum multiplexer: (a) Detailed circuit; (b) Simplified circuit.}
			\label{F2}
		\end{figure}

		In classical logic, there is an if-then-else structure. In quantum computing, the corresponding concept is the quantum multiplexer \citep{R29}, and its unitary matrix is generally as follows:
		\begin{equation}
			\mathbf{U}=\underset{k=0}{\overset{N-1}{\mathop{\oplus }}}\,{{\mathbf{T}}_{k}}=\left[ \begin{matrix}
				{{\mathbf{T}}_{0}} & {} & {} & {}  \\
				{} & {{\mathbf{T}}_{1}} & {} & {}  \\
				{} & {} & \ddots  & {}  \\
				{} & {} & {} & {{\mathbf{T}}_{N-1}}  \\
			\end{matrix} \right],
			\label{E6}
		\end{equation}
		whose quantum circuit is shown in Fig. \ref{F2} (a). Through the circuit, a series of conditional operations can be realized: when the quantum state of the control qubit is $\left| j \right\rangle $, the operation ${{\mathbf{T}}_{j}}$ is applied to the target qubit. For convenience of discussion, the quantum circuit of this operation is uniformly represented as the form shown in Fig. \ref{F2} (b).
		
		It can be seen that the quantum multiplexer consists of $N$ multi-qubit control gates, each operating on $n={{\log }_{2}}N$ qubits, so the implementation of the quantum multiplexer is relatively complex. Currently, two-qubit control gates are gradually maturing, and hardware research on three-qubit control gates is also underway \citep{R30}. Therefore, when multi-qubit control gates cannot be directly implemented on current quantum computers, many scholars have studied decomposing them into multiple two-qubit control gates. For example, literature \citep{R31,R32} introduces methods to realize multi-qubit control gates using $\text{poly}\left( n \right)$ two-qubit control gates. Since a quantum multiplexer is composed of $N$ multi-qubit control gates, it requires $\text{poly}\left( n \right)N$ gates, whose computational efficiency fails to reflect the advantages of quantum computing. Quantum random access memory (QRAM) is a potential way to implement quantum multiplexers. QRAM leverages quantum parallelism and can perform the task with only $\text{poly}\left( {{\log }_{2}}N \right)$ operations. Although QRAM has not been fully and effectively realized physically yet, an increasing number of studies on QRAM are emerging \citep{R33,R34,R35,R36}. It has already been possible to physically implement the two-qubit quantum multiplexer shown in Fig. 1(e) and multi-qubit control gates using neutral atoms \citep{R37,R38,R39}. Therefore, it is believable that quantum multiplexers can be effectively realized with $\text{poly}\left( n \right)$ complexity in the future. It must be pointed out that quantum multiplexers are key steps in many quantum algorithms such as the HHL algorithm, quantum fluid dynamics simulation, quantum data mining, quantum regression, and so on \citep{R4,R40,R41,R42,R43}. The aforementioned research assumes that quantum multiplexers can be realized with $\text{poly}\left( n \right)$ complexity. Thus, this study is also based on the premise that quantum multiplexers can be effectively realized with $\text{poly}\left( n \right)$ complexity.
		
		\section{Structure and Decomposition of Matrices Generated by Voxel Grids}
		\label{S3}
		The voxel grid is one of the earliest discretization methods used in solving differential equations and has extensive successful applications in solid mechanics problems. The voxel grid is characterized by dividing the solution domain with regular grids. Compared with traditional triangular grids, the voxel grid offers the advantage of simple modeling, though it suffers from high time and space complexity. Recently, Guo et al. \citep{R25,R26,R27} combined machine learning with computational mechanics and proposed the problem-independent machine learning method, achieving two orders of magnitude higher computational efficiency than traditional finite elements. Their work enables efficient analysis of finite elements under ultra-large-scale voxel grids, indicating that with appropriately designed computational frameworks, efficient and high-precision finite element simulations using large-scale voxel grids are achievable. The inherent advantages of quantum computers in parallel computing exhibit excellent compatibility with voxel grids. Developing quantum computational mechanics methods based on voxel grids can be expected to achieve efficient and real-time mechanical simulations. This section analyzes the characteristics of matrices generated during voxel grid modeling and further discusses the decomposition of the generated matrices.
		
		\subsection{Fractal Property of Matrices Generated by Voxel Grids}
		\label{S3S1}
\begin{figure}[!h]
			\centering
			\includegraphics[width=1\textwidth]{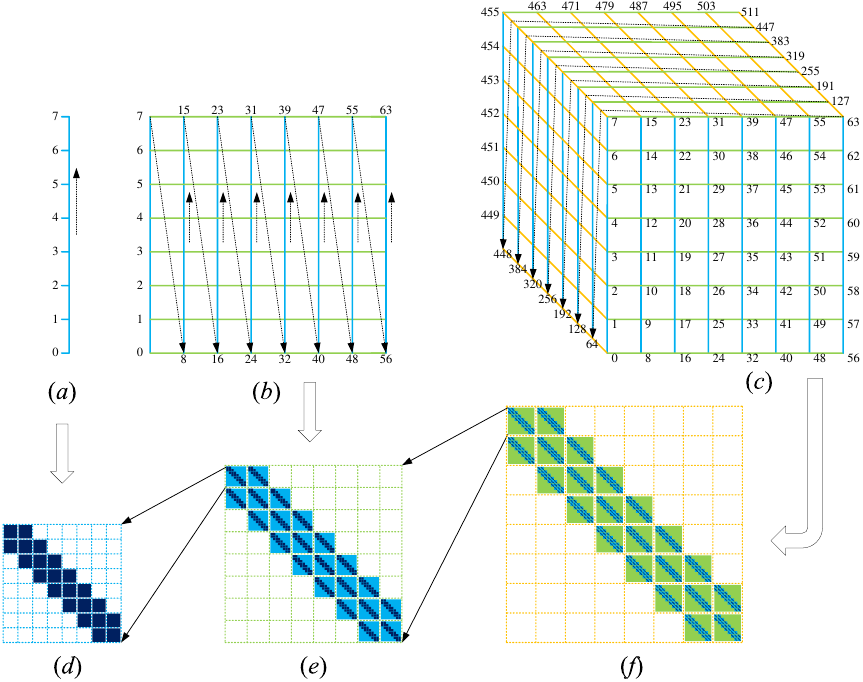}
			\caption{The fractal property of the matrices generated by the voxel grid: (a) Schematic of a 1-dimensional voxel grid with ${{N}_{y}}=8$ nodes; (b) Schematic of a 2-dimensional voxel grid with ${{N}_{x}}=8$, ${{N}_{y}}=8$, and a total of 64 nodes; (c) Schematic of a 3-dimensional voxel grid with ${{N}_{x}}=8$, ${{N}_{y}}=8$, ${{N}_{z}}=8$, and a total of 512 nodes; (d) Stiffness matrix generated by the 1-dimensional voxel grid in (a), where dark blue regions represent d-dimensional submatrices and white regions are zero; (e) Stiffness matrix generated by the 2-dimensional voxel grid in (b), where light blue regions represent submatrices corresponding to the 1-dimensional voxel grid and white regions are zero; (f) Stiffness matrix generated by the 3-dimensional voxel grid in (c), where green regions represent submatrices corresponding to the 2-dimensional voxel grid and white regions are zero.}
			\label{F3}
		\end{figure}

		The voxel grids discretize the domain using relatively regular grids, as shown in Fig. \ref{F3}, which illustrates voxel grid discretizations for 1-dimensional, 2-dimensional, and 3-dimensional regular problems. The node numbering order is closely related to the structure of the generated matrix. The simplest case is the 1-dimensional voxel grid, as shown in Fig. \ref{F3} (a), which contains ${{N}_{y}}$ nodes numbered sequentially from bottom to top. For readers familiar with finite element or finite difference methods, the numbering results in a tridiagonal block matrix which is shown in Fig. \ref{F3} (d), where the dimension of each block submatrix equals the number of variables per node. It can be seen from Fig. \ref{F3} (b) that the 2-dimensional voxel grid can be viewed as consisting of multiple 1-dimensional sub-blocks voxel grids, with each 1-dimensional node also numbered in ascending order. Under the numbering, the resulting matrix is also a tridiagonal block matrix, where each sub-block corresponds to a matrix generated by a 1-dimensional voxel grid so as each sub-block matrix is also tridiagonal which can be seen in Fig. \ref{F3} (e). Similarly, a 3-dimensional voxel grid is composed of multiple 2-dimensional sub-block grids, as shown in Fig. \ref{F3} (c), with node numbering grouped by 2-dimensional sub-blocks. Each 2-dimensional sub-block, in turn, is composed of 1-dimensional sub-blocks, leading to a larger tridiagonal block matrix where each sub-block represents a matrix generated by the 2-dimensional voxel grid, which can be seen in Fig. \ref{F3} (f).
		
		According to the node numbering properties of voxel grids, it is evident that MVGs exhibit fractal characteristics. The 1-dimensional MVG is a fractal substructure of the 2-dimensional MVG, while the 2-dimensional MVG is also a fractal substructure of a 3-dimensional MVG. It is shown in Figs. \ref{F3} (d)-(e) that each basic fractal component is a tridiagonal block matrix. Leveraging the fractal structure, the matrix generated by the higher-dimensional voxel grids can be easily derived, which is not elaborated here for brevity.
		
		\subsection{Special Cases}
		\label{S3S2}
		Subsection \ref{S3S1} introduced regular voxel grids and the corresponding matrices. This subsection will discuss the handling of special cases, focusing on two scenarios: irregular solution domains, and special treatments for the number of node variables.
		
		\subsubsection{Irregular Regions}
		\label{S3S2S1}
		Irregular domains can also be discretized by voxel grids. Here, two approaches will be introduced with a 2-dimensional irregular shape as an example, with straightforward generalization to multidimensional cases.
		
		\begin{figure}[!h]
			\centering
			\includegraphics[width=1\textwidth]{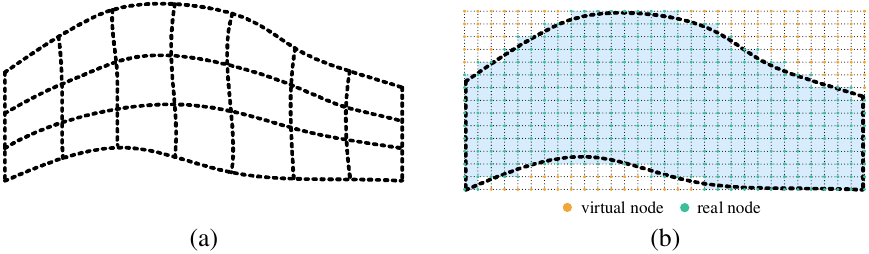}
			\caption{Voxel grids for irregular domains: (a) Homeomorphic voxel grid; (b) Regular voxel grid.}
			\label{F4}
		\end{figure}
		
		For the irregular domains shown in Fig. \ref{F4}, two types of grids can be used for discretization. The first one is characterized by being homeomorphic to the voxel grid of regular domains, as shown in Fig. \ref{F4} (a). Therefore, the matrices generated by the grids also exhibit the tridiagonal fractal property of the MVG. The second type still uses a regular voxel grid to discretize the irregular domains. When the number of voxel grids is sufficiently large, it can accurately approximate the irregular domain, as shown in Fig. \ref{F4} (b). For the grid in Fig. \ref{F4} (b), the concept of virtual nodes is introduced: when a node is outside the solution domain, its variables are still treated as virtual variables to be solved. In the corresponding matrix, zero elements are filled in the corresponding positions, and a constant $C$ is placed on the diagonal. The resulting matrix still maintains the tridiagonal block structure of the matrix generated by the voxel grid as shown in Fig. \ref{F3}. Although the regular voxel grid approach introduces virtual variables for irregular domains, it is highly suitable for implementation on quantum computers. In quantum computers, the number of qubits generally needs to be $\left\lceil {{\log }_{2}}\left( Nd \right) \right\rceil $, where $N$ is the total number of nodes and $d$ is the number of variables per node. Discretizing each direction into ${{2}^{{{n}_{x}}}}\times {{2}^{{{n}_{y}}}}\times {{2}^{{{n}_{z}}}}$ for irregular domains using regular voxel grids aligns with the physical properties of quantum computers, which naturally support binary-structured computations.
		
		\subsubsection{Influence of the Number of Node Variables}
		\label{S3S2S2}
\begin{figure}[!h]
			\centering
			\includegraphics[width=1\textwidth]{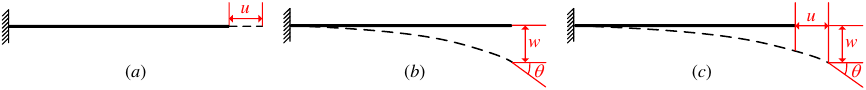}
			\caption{Deformation of a one-dimensional rod: (a) Single-bar tension problem, single-variable; (b) Pure bending problem, two-variable; (c) Combined tension and bending problem, three-variable.}
			\label{F5}
		\end{figure}

		Common mechanical analyses, such as elasticity problems, heat conduction and diffusion problems, and wave propagation problems, can be classified into single-variable problems, two-variable problems, and three-variable problems based on the number of unknowns to be solved. Taking a one-dimensional rod as an example, as shown in Fig. \ref{F5}: when considering the tension and compression problems of the rod, it is a single-variable problem; when considering the bending of the rod, it becomes a two-variable problem, where the unknown variables typically include deflection and rotation angle; when considering both the bending and tension of the rod, it becomes a three-variable problem, with unknown variables including not only vertical deflection but also lateral displacement and rotation angle. For some special mechanical problems, such as the bending of a rod with piezoelectric coupling, even more types of variables may be involved. Due to space limitations, this study mainly focuses on one-variable, two-variable, and three-variable problems to construct corresponding Hamiltonian simulation algorithms. The proposed quantum algorithms can be easily extended to multi-variable problems.
				
		Since the number of variables that a quantum computer can handle is generally ${{2}^{n}}$, when using voxel grids for discretization, the number of nodes in each direction is chosen as a power of 2. In this case, for a problem with $d$ variables, the total degrees of freedom of the structure are ${{2}^{n}}\times d$. Notably, when $d=3$, ${{2}^{n}}\times d$ is not a power of 2. To address it, the concept of virtual nodes is reintroduced. Taking the combined bending and tension problem of a beam as an example for studying the quantum simulation of three-variable problems, suppose the beam is discretized into ${{N}_{x}}$ nodes along the length, with each node having three variables. The overall displacement vector can be written as,
		\begin{equation}
			\mathbf{u}={{\left( \begin{matrix}
						{{\mathbf{u}}_{0}} & {{\mathbf{u}}_{1}} & \cdots  & {{\mathbf{u}}_{{{N}_{x}}-1}}  \\
					\end{matrix} \right)}^{\text{T}}},\ \ \ {{\mathbf{u}}_{i}}={{\left( {{u}_{i}},\ {{w}_{i}},\ {{\theta }_{i}} \right)}^{\text{T}}}.
				\label{E7}
		\end{equation}
		When performing numerical calculations for the problem, if ${{N}_{x}}={{2}^{{{n}_{x}}}}$ nodes are taken, there will be a total of $3\times {{2}^{{{n}_{x}}}}$ variables. Therefore, at least $\left\lceil {{\log }_{2}}\left( 3\times {{2}^{{{n}_{x}}}} \right) \right\rceil =\left\lceil {{\log }_{2}}3 \right\rceil +{{n}_{x}}={{n}_{x}}+2$ qubits are required. This is equivalent to assigning 4 variables per node, where the first three variables are $\left( {{u}_{i}},\ {{w}_{i}},\ {{\theta }_{i}} \right)$, and the fourth is an unnecessary virtual variable. When solving the problem, a tridiagonal block matrix as shown in Eq. (\ref{E8}) can also be generated,
		\begin{equation}
			\mathbf{K}=\left[ \begin{matrix}
				{{\mathbf{k}}_{00}} & {{\mathbf{k}}_{01}} & {} & {} & {}  \\
				{{\mathbf{k}}_{10}} & {{\mathbf{k}}_{11}} & {{\mathbf{k}}_{12}} & {} & {}  \\
				{} & {{\mathbf{k}}_{21}} & {{\mathbf{k}}_{22}} & \ddots  & {}  \\
				{} & {} & \ddots  & \ddots  & {{\mathbf{k}}_{{{N}_{x}}-2,{{N}_{x}}-1}}  \\
				{} & {} & {} & {{\mathbf{k}}_{{{N}_{x}}-1,{{N}_{x}}-2}} & {{\mathbf{k}}_{{{N}_{x}}-1,{{N}_{x}}-1}}  \\
			\end{matrix} \right],
			\label{E8}
		\end{equation}
		where ${{\mathbf{k}}_{ij}}$ is a $4\times 4$ matrix including elements corresponding to a virtual node, in the form of:
		\begin{equation}
			{{\mathbf{k}}_{ij}}=\left\{ \begin{aligned}
				& {{{\mathbf{\hat{k}}}}_{ii}}\oplus C,\ \ i=j, \\ 
				& {{{\mathbf{\hat{k}}}}_{ij}}\oplus 0,\ \ \ i\ne j, \\ 
			\end{aligned} \right.
		\end{equation}
		where $C$ is an arbitrary constant (e.g., set to 1), and ${{\mathbf{\hat{k}}}_{ij}}$ is a $3\times 3$ matrix obtained through numerical methods such as the finite element method.
		
		\subsection{Decomposition of Matrices Generated by Voxel Grids}
		\label{S3S3}
		Subsections \ref{S3S1} and \ref{S3S2} have investigated the characteristics of the MVG. In quantum computing, calculating the matrix exponential is a very crucial step, and the core difficulty lies in how to decompose the matrix. To suit the characteristics of quantum computers, this section leverages the tridiagonal fractal property of the MVG and combines the direct product and direct sum operations of matrices to present the KCQ decomposition method for the MVG.
			
		\subsubsection{KCQ decomposition of the system matrices generated by 1-dimensional voxel grids}
		\label{S3S3S1}
		Let the 1-dimensional problem be discretized along the $x$ direction with ${{N}_{x}}={{2}^{{{n}_{x}}}}$ nodes. The corresponding MVG is shown in Eq. (\ref{E8}), where ${{\mathbf{k}}_{ij}}$ is a ${{2}^{\left\lceil {{\log }_{2}}d \right\rceil }}\text{-dimensional}$ sub-block matrix. To compute the matrix exponential ${{e}^{\text{i}\Delta t\mathbf{K}}}$, it is necessary to decompose matrix $\mathbf{K}$	into a finite number of matrices ${{\mathbf{K}}_{k}}$ (which need to be unitary or Hermitian matrices), and the matrix exponential of each ${{\mathbf{K}}_{k}}$ can be effectively simulated by quantum computing. In general, the decomposition work is very challenging. However, under the voxel grid framework, it is found that an effective decomposition of $\mathbf{K}$	can be achieved using only five matrices ${{\mathbf{K}}_{k}}$, $k=1,\ \cdots ,\ 5$, each of which is a Hermitian matrix and whose matrix exponential can be easily computed. Let
		\begin{equation}
			\mathbf{K}=\mathbf{K}_{1}^{{}}+\mathbf{K}_{2}^{{}}+\mathbf{K}_{3}^{{}}+\mathbf{K}_{4}^{{}}+\mathbf{K}_{5}^{{}},
			\label{E10}
		\end{equation}
		where
		\begin{equation}
			\begin{aligned}
				& {\mathbf{K}_{1}^{{}}=\left[ \begin{matrix}
						{{\mathbf{k}}_{00}} & {} & {} & {}  \\
						{} & {{\mathbf{k}}_{11}} & {} & {}  \\
						{} & {} & \ddots  & {}  \\
						{} & {} & {} & {{\mathbf{k}}_{{{N}_{x}}-1,{{N}_{x}}-1}}  \\
					\end{matrix} \right],} \\
				\end{aligned}
			\label{E11}
		\end{equation}
				\begin{equation}
			\begin{aligned}	
				& {\mathbf{K}_{2}^{{}}=\left[ \begin{matrix}
						{} & {{\mathbf{c}}_{0}} & {} & {} & {} & {} & {}  \\
						{{\mathbf{c}}_{0}} & {} & \mathbf{0} & {} & {} & {} & {}  \\
						{} & \mathbf{0} & {} & {{\mathbf{c}}_{2}} & {} & {} & {}  \\
						{} & {} & {{\mathbf{c}}_{2}} & {} & \mathbf{0} & {} & {}  \\
						{} & {} & {} & \mathbf{0} & {} & {{\mathbf{c}}_{4}} & {}  \\
						{} & {} & {} & {} & {{\mathbf{c}}_{4}} & {} & \ddots   \\
						{} & {} & {} & {} & {} & \ddots  & {}  \\
					\end{matrix} \right],} \\
				& {\mathbf{K}_{3}^{{}}=\left[ \begin{matrix}
						{} & -\text{i}{{\mathbf{q}}_{0}} & {} & {} & {} & {} & {}  \\
						\text{i}{{\mathbf{q}}_{0}} & {} & \mathbf{0} & {} & {} & {} & {}  \\
						{} & \mathbf{0} & {} & -\text{i}{{\mathbf{q}}_{2}} & {} & {} & {}  \\
						{} & {} & \text{i}{{\mathbf{q}}_{2}} & {} & \mathbf{0} & {} & {}  \\
						{} & {} & {} & \mathbf{0} & {} & -\text{i}{{\mathbf{q}}_{4}} & {}  \\
						{} & {} & {} & {} & \text{i}{{\mathbf{q}}_{4}} & {} & \ddots   \\
						{} & {} & {} & {} & {} & \ddots  & {}  \\
					\end{matrix} \right],} \\									
				& {\mathbf{K}_{4}^{{}}=\left[ \begin{matrix}
						{} & \mathbf{0} & {} & {} & {} & {} & {}  \\
						\mathbf{0} & {} & {{\mathbf{c}}_{1}} & {} & {} & {} & {}  \\
						{} & {{\mathbf{c}}_{1}} & {} & \mathbf{0} & {} & {} & {}  \\
						{} & {} & \mathbf{0} & {} & {{\mathbf{c}}_{3}} & {} & {}  \\
						{} & {} & {} & {{\mathbf{c}}_{3}} & {} & \mathbf{0} & {}  \\
						{} & {} & {} & {} & \mathbf{0} & {} & \ddots   \\
						{} & {} & {} & {} & {} & \ddots  & {}  \\
					\end{matrix} \right],} \\		
				& {\mathbf{K}_{5}^{{}}=\left[ \begin{matrix}
						{} & \mathbf{0} & {} & {} & {} & {} & {}  \\
						\mathbf{0} & {} & -\text{i}{{\mathbf{q}}_{1}} & {} & {} & {} & {}  \\
						{} & \text{i}{{\mathbf{q}}_{1}} & {} & \mathbf{0} & {} & {} & {}  \\
						{} & {} & \mathbf{0} & {} & -\text{i}{{\mathbf{q}}_{3}} & {} & {}  \\
						{} & {} & {} & \text{i}{{\mathbf{q}}_{3}} & {} & \mathbf{0} & {}  \\
						{} & {} & {} & {} & \mathbf{0} & {} & \ddots   \\
						{} & {} & {} & {} & {} & \ddots  & {}  \\
					\end{matrix} \right],} 
			\end{aligned}
			\label{E12}
		\end{equation}
		\begin{equation}
			{{\mathbf{c}}_{n}}\left( \mathbf{K} \right)=\frac{1}{2}\left( {{\mathbf{k}}_{n,n+1}}+{{\mathbf{k}}_{n+1,n}} \right),\ \ \ {{\mathbf{q}}_{n}}\left( \mathbf{K} \right)=\frac{\text{i}}{2}\left( {{\mathbf{k}}_{n,n+1}}-{{\mathbf{k}}_{n+1,n}} \right).
			\label{E13}
		\end{equation}
		In the above decomposition, direct computation of ${{\mathbf{K}}_{4}}$ and ${{\mathbf{K}}_{5}}$ is not easily achievable. However, by leveraging the properties of the direct product $\otimes $ and direct sum $\oplus $ of matrices, along with the Pauli-X gate $\mathbf{X}$, Pauli-Y gate $\mathbf{Y}$ in quantum computing, and circulant matrix ${{\mathbf{S}}_{{{N}_{x}}}}=\left[ \begin{matrix}
			\mathbf{0} & {{\mathbf{I}}_{{{N}_{x}}-1}}  \\
			1 & \mathbf{0}  \\
		\end{matrix} \right]$, Eqs. (\ref{E11}) and (\ref{E12}) can be further expressed as
		\begin{equation}
			\begin{aligned}
				& {{\mathbf{K}}_{1}}=\underset{n=0}{\overset{{{2}^{{{n}_{x}}}}-1}{\mathop{\oplus }}}\,{{\mathbf{k}}_{nn}},\ \ {{\mathbf{K}}_{2}}=\underset{n=0}{\overset{{{2}^{{{n}_{x}}-1}}-1}{\mathop{\oplus }}}\,\left( \mathbf{X}\otimes {{\mathbf{c}}_{2n}} \right),\ \ {{\mathbf{K}}_{3}}=\underset{n=0}{\overset{{{2}^{{{n}_{x}}-1}}-1}{\mathop{\oplus }}}\,\left( \mathbf{Y}\otimes {{\mathbf{q}}_{2n}} \right), \\ 
				& {{\mathbf{K}}_{4}}=\left( \mathbf{S}_{{{N}_{x}}}^{\text{H}}\otimes {{\mathbf{I}}_{d}} \right)\left[ \underset{n=0}{\overset{{{2}^{{{n}_{x}}-1}}-1}{\mathop{\oplus }}}\,\left( \mathbf{X}\otimes {{\mathbf{c}}_{2n+1}} \right) \right]\left( {{\mathbf{S}}_{{{N}_{x}}}}\otimes {{\mathbf{I}}_{d}} \right), \\ 
				& {{\mathbf{K}}_{5}}=\left( \mathbf{S}_{{{N}_{x}}}^{\text{H}}\otimes {{\mathbf{I}}_{d}} \right)\left[ \underset{n=0}{\overset{{{2}^{{{n}_{x}}-1}}-1}{\mathop{\oplus }}}\,\left( \mathbf{Y}\otimes {{\mathbf{q}}_{2n+1}} \right) \right]\left( {{\mathbf{S}}_{{{N}_{x}}}}\otimes {{\mathbf{I}}_{d}} \right), \\ 
			\end{aligned}
			\label{E14}
		\end{equation}	
		where ${{\mathbf{c}}_{{{2}^{{{n}_{x}}}}-1}}={{\mathbf{q}}_{{{2}^{{{n}_{x}}}}-1}}=\mathbf{0}$. Since the Pauli-X gate $\mathbf{X}$ and Pauli-Y gate $\mathbf{Y}$ are not only unitary matrices but also Hermitian matrices, they can be diagonalized via the eigenvalues and eigenvectors as
		\begin{equation}
			\mathbf{X}=\mathbf{H}\left[ \begin{matrix}
				1 & {}  \\
				{} & -1  \\
			\end{matrix} \right]\mathbf{H},\ \ \ \mathbf{Y}=\mathbf{T}\left( 0.5\pi  \right)\mathbf{H}\left[ \begin{matrix}
				1 & {}  \\
				{} & -1  \\
			\end{matrix} \right]\mathbf{HT}\left( 1.5\pi  \right),
			\label{E15}
		\end{equation}
		where $\mathbf{H}$ and $\mathbf{T}$ represent the Hadamard gate and the phase gate, respectively. Substituting Eq. (\ref{E15}) into Eq. (\ref{E14}) yields
		\begin{equation}
			\begin{aligned}
				& {{\mathbf{K}}_{1}}=\underset{n=0}{\overset{{{2}^{{{n}_{x}}}}-1}{\mathop{\oplus }}}\,{{\mathbf{k}}_{nn}},\ \ \  \\ 
				& {{\mathbf{K}}_{2}}={{\mathbf{L}}_{2,{{N}_{x}},d}}\left[ \underset{n=0}{\overset{{{2}^{{{n}_{x}}}}-1}{\mathop{\oplus }}}\,{{\left( -1 \right)}^{n}}{{\mathbf{c}}_{2\left\lfloor 0.5n \right\rfloor }} \right]\mathbf{L}_{2,{{N}_{x}},d}^{\text{H}} \\ 
				& {{\mathbf{K}}_{3}}={{\mathbf{L}}_{3,{{N}_{x}},d}}\left[ \underset{n=0}{\overset{{{2}^{{{n}_{x}}}}-1}{\mathop{\oplus }}}\,{{\left( -1 \right)}^{n}}{{\mathbf{q}}_{2\left\lfloor 0.5n \right\rfloor }} \right]\mathbf{L}_{3,{{N}_{x}},d}^{\text{H}} \\ 
				& {{\mathbf{K}}_{4}}={{\mathbf{L}}_{4,{{N}_{x}},d}}\left[ \underset{n=0}{\overset{{{2}^{{{n}_{x}}}}-1}{\mathop{\oplus }}}\,{{\left( -1 \right)}^{n}}{{\mathbf{c}}_{2\left\lfloor 0.5n \right\rfloor +1}} \right]\mathbf{L}_{4,{{N}_{x}},d}^{\text{H}} \\ 
				& {{\mathbf{K}}_{5}}={{\mathbf{L}}_{5,{{N}_{x}},d}}\left[ \underset{n=0}{\overset{{{2}^{{{n}_{x}}}}-1}{\mathop{\oplus }}}\,{{\left( -1 \right)}^{n}}{{\mathbf{q}}_{2\left\lfloor 0.5n \right\rfloor +1}} \right]\mathbf{L}_{5,{{N}_{x}},d}^{\text{H}} \\ 
			\end{aligned},
			\label{E16}
		\end{equation}
		where
		\begin{equation}
			\begin{aligned}
				& {{\mathbf{L}}_{2,{{N}_{x}},d}}={{\mathbf{I}}_{{{2}^{{{n}_{x}}-1}}}}\otimes \mathbf{H}\otimes {{\mathbf{I}}_{d}},\ \ \  \\ 
				& {{\mathbf{L}}_{3,{{N}_{x}},d}}={{\mathbf{I}}_{{{2}^{{{n}_{x}}-1}}}}\otimes \left( \mathbf{T}\left( 0.5\pi  \right)\mathbf{H} \right)\otimes {{\mathbf{I}}_{d}},\ \ \  \\ 
				& {{\mathbf{L}}_{4,{{N}_{x}},d}}=\left( \mathbf{S}_{{{N}_{x}}}^{\text{H}}\otimes {{\mathbf{I}}_{d}} \right)\left( {{\mathbf{I}}_{{{2}^{{{n}_{x}}-1}}}}\otimes \mathbf{H}\otimes {{\mathbf{I}}_{d}} \right), \\ 
				& {{\mathbf{L}}_{5,{{N}_{x}},d}}=\left( \mathbf{S}_{{{N}_{x}}}^{\text{H}}\otimes {{\mathbf{I}}_{d}} \right)\left[ {{\mathbf{I}}_{{{2}^{{{n}_{x}}-1}}}}\otimes \left( \mathbf{T}\left( 0.5\pi  \right)\mathbf{H} \right)\otimes {{\mathbf{I}}_{d}} \right], 
			\end{aligned}
			\label{E17}
		\end{equation}
		and $d$ is related to the problem. When dealing with single-variable, two-variable, or three-variable problems, $d$ is 1, 2, or 4, respectively. Through matrix decomposition, Eq. (\ref{E16}) expresses $\mathbf{K}$ as five Hermitian matrices with ${{\mathbf{k}}_{nn}}$, ${{\mathbf{c}}_{n}}$, and ${{\mathbf{q}}_{n}}$ as the basic elements, which can be represented by basic quantum gates. Since the proposed method decomposes $\mathbf{K}$ into ${{\mathbf{k}}_{nn}}$, ${{\mathbf{c}}_{n}}$, and ${{\mathbf{q}}_{n}}$, the proposed decomposition method is simply referred to as the KCQ decomposition in this paper. When dealing with a single-variable problem, ${{\mathbf{q}}_{n}}=0$, ${{\mathbf{k}}_{nn}}$ and ${{\mathbf{c}}_{n}}$ are real numbers. When dealing with a two-variable problem (or three-variable problem), ${{\mathbf{k}}_{nn}}$, ${{\mathbf{c}}_{n}}$ and ${{\mathbf{q}}_{n}}$ are 2-dimensional Hermitian matrices (or 4-dimensional Hermitian matrices), and the matrix exponentials can be easily represented by basic quantum gates. Therefore, by combining the decomposition shown in Eq. (\ref{E16}) with Lemma 1 (see \ref{APPA}), the effective calculation of ${{e}^{\text{i}\Delta t\mathbf{K}}}$ can be easily achieved. The specific calculation scheme and the corresponding quantum circuit are presented in Section \ref{S4}.
		
		\subsubsection{Recursive KCQ Decomposition of Matrices Generated by Multidimensional Voxel Grid}
		\label{S3S3S2}
		Equation (\ref{E10}) presents the decomposition of the system matrices generated by the 1-dimensional voxel grids. For the system matrices generated by multi-dimensional voxel grids, the KCQ decomposition shown in Eq. (\ref{E10}) remains applicable. Taking the 2-dimensional voxel grid as an example, according to Fig. 3, the generated matrix is also a tridiagonal block matrix with a fractal structure. That is, each sub-block matrix within the 2-dimensional MVG is still a tridiagonal block matrix. Therefore, the KCQ decomposition can still be used for the system matrices generated by the 2-dimensional voxel grids. If the two-dimensional problem is discretized with ${{2}^{{{n}_{x}}}}$ and ${{2}^{{{n}_{y}}}}$ nodes along the x and y directions respectively, using the KCQ decomposition, the 2-dimensional MVG can be decomposed into
		\begin{equation}
			\begin{aligned}
				 \mathbf{K}&=\underset{n=0}{\overset{{{2}^{{{n}_{x}}}}-1}{\mathop{\oplus }}}\,{{\mathbf{k}}_{nn}}\left( \mathbf{K} \right) \\ 
				&+{{\mathbf{L}}_{2,{{N}_{x}},{{N}_{y}}d}}\left[ \underset{n=0}{\overset{{{2}^{{{n}_{x}}}}-1}{\mathop{\oplus }}}\,{{\left( -1 \right)}^{n}}{{\mathbf{c}}_{2\left\lfloor 0.5n \right\rfloor }}\left( \mathbf{K} \right) \right]\mathbf{L}_{2,{{N}_{x}},{{N}_{y}}d}^{\text{H}}\\
				&+{{\mathbf{L}}_{3,{{N}_{x}},{{N}_{y}}d}}\left[ \underset{n=0}{\overset{{{2}^{{{n}_{x}}}}-1}{\mathop{\oplus }}}\,{{\left( -1 \right)}^{n}}{{\mathbf{q}}_{2\left\lfloor 0.5n \right\rfloor }}\left( \mathbf{K} \right) \right]\mathbf{L}_{3,{{N}_{x}},{{N}_{y}}d}^{\text{H}} \\ 
				&+{{\mathbf{L}}_{4,{{N}_{x}},{{N}_{y}}d}}\left[ \underset{n=0}{\overset{{{2}^{{{n}_{x}}}}-1}{\mathop{\oplus }}}\,{{\left( -1 \right)}^{n}}{{\mathbf{c}}_{2\left\lfloor 0.5n \right\rfloor +1}}\left( \mathbf{K} \right) \right]\mathbf{L}_{4,{{N}_{x}},{{N}_{y}}d}^{\text{H}}\\
				&+{{\mathbf{L}}_{5,{{N}_{x}},{{N}_{y}}d}}\left[ \underset{n=0}{\overset{{{2}^{{{n}_{x}}}}-1}{\mathop{\oplus }}}\,{{\left( -1 \right)}^{n}}{{\mathbf{q}}_{2\left\lfloor 0.5n \right\rfloor +1}}\left( \mathbf{K} \right) \right]\mathbf{L}_{5,{{N}_{x}},{{N}_{y}}d}^{\text{H}}
			\end{aligned},
			\label{E18}
		\end{equation}
		where the expressions for ${{\mathbf{k}}_{nn}}\left( \mathbf{K} \right),\ {{\mathbf{c}}_{n}}\left( \mathbf{K} \right),\ {{\mathbf{q}}_{n}}\left( \mathbf{K} \right)$ are shown in Eqs. (\ref{E11})-(\ref{E13}). For the convenience of discussion, the three matrices are respectively labeled as $\mathbf{k}_{nn}^{\left( 1 \right)}$, $\mathbf{c}_{n}^{\left( 1 \right)}$ and $\mathbf{q}_{n}^{\left( 1 \right)}$. Obviously, these three matrices are all matrices generated by 1-dimensional voxel grids, thus tridiagonal matrices in the form of Eq. (\ref{E8}). Therefore, we can continue to apply the KCQ decomposition to $\mathbf{k}_{nn}^{\left( 1 \right)}$, $\mathbf{c}_{n}^{\left( 1 \right)}$, and $\mathbf{q}_{n}^{\left( 1 \right)}$. Taking $\mathbf{k}_{nn}^{\left( 1 \right)}$ as an example, the KCQ decomposition is as follows,
		\begin{equation}
			\begin{aligned}
				\mathbf{k}_{nn}^{\left( 1 \right)}&=\underset{m=0}{\overset{{{2}^{{{n}_{y}}}}-1}{\mathop{\oplus }}}\,{{\mathbf{k}}_{mm}}\left( \mathbf{k}_{nn}^{\left( 1 \right)} \right) \\ 
				&+{{\mathbf{L}}_{2,{{N}_{y}},d}}\left[ \underset{m=0}{\overset{{{2}^{{{n}_{y}}}}-1}{\mathop{\oplus }}}\,{{\left( -1 \right)}^{m}}{{\mathbf{c}}_{2\left\lfloor 0.5m \right\rfloor }}\left( \mathbf{k}_{nn}^{\left( 1 \right)} \right) \right]\mathbf{L}_{2,{{N}_{y}},d}^{\text{H}}\\
				&+{{\mathbf{L}}_{3,{{N}_{y}},d}}\left[ \underset{m=0}{\overset{{{2}^{{{n}_{y}}}}-1}{\mathop{\oplus }}}\,{{\left( -1 \right)}^{m}}{{\mathbf{q}}_{2\left\lfloor 0.m \right\rfloor }}\left( \mathbf{k}_{nn}^{\left( 1 \right)} \right) \right]\mathbf{L}_{3,{{N}_{y}},d}^{\text{H}} \\ 
				&+{{\mathbf{L}}_{4,{{N}_{y}},d}}\left[ \underset{m=0}{\overset{{{2}^{{{n}_{y}}}}-1}{\mathop{\oplus }}}\,{{\left( -1 \right)}^{m}}{{\mathbf{c}}_{2\left\lfloor 0.5m \right\rfloor +1}}\left( \mathbf{k}_{nn}^{\left( 1 \right)} \right) \right]\mathbf{L}_{4,{{N}_{y}},d}^{\text{H}}\\
				&+{{\mathbf{L}}_{5,{{N}_{y}},d}}\left[ \underset{m=0}{\overset{{{2}^{{{n}_{y}}}}-1}{\mathop{\oplus }}}\,{{\left( -1 \right)}^{m}}{{\mathbf{q}}_{2\left\lfloor 0.5m \right\rfloor +1}}\left( \mathbf{k}_{nn}^{\left( 1 \right)} \right) \right]\mathbf{L}_{5,{{N}_{y}},d}^{\text{H}} 
			\end{aligned},
			\label{E19}
		\end{equation}
		where the matrix ${{\mathbf{k}}_{mm}}\left( \mathbf{k}_{nn}^{\left( 1 \right)} \right),\ {{\mathbf{c}}_{m}}\left( \mathbf{k}_{nn}^{\left( 1 \right)} \right),\ {{\mathbf{q}}_{m}}\left( \mathbf{k}_{nn}^{\left( 1 \right)} \right)$ are already $d\text{-dimensional}$ ($d=1,\ 2,\ 4$) matrices, whose matrix exponentials can be easily represented by basic quantum gates.
		
		For the decomposition of the matrices generated by higher-dimensional voxel grids, the proposed KCQ decomposition method can still be used during the recursive iteration. Each iteration can reduce the dimension of the matrices through KCQ decomposition, eventually resulting in basic $\mathbf{k}$, $\mathbf{c}$, and $\mathbf{q}$ matrices with only d dimensions. Due to space constraints, the decomposition details are not elaborated here.
		
		\section{VBQC for Hamiltonian Simulation Based on KCQ Decomposition}
		\label{S4}
		This section provides a detailed description of the VBQC and the corresponding quantum circuit for the QSH in general solid mechanics. For the convenience of illustration, we assign two marker symbols for the quantum circuits. Suppose $\mathbf{A}$ is a $d\text{-dimensional}$ Hermitian matrix,
		\begin{equation}
			\mathbf{A}=\left[ \begin{matrix}
				{{\mathbf{a}}_{11}} & {{\mathbf{a}}_{12}}  \\
				{{\mathbf{a}}_{21}} & {{\mathbf{a}}_{22}}  \\
			\end{matrix} \right],
			\label{E20}
		\end{equation}
		and the quantum circuit for the matrix exponential ${{e}^{\text{i}\Delta t\mathbf{A}}}$ is denoted as ${{\mathbf{W}}_{d}}\left( \mathbf{A}\Delta t \right)$.
		
		The Hamiltonian $\mathbf{K}$ under the $m\text{-dimensional}$ voxel grids represents a tridiagonal block matrix satisfying the fractal property shown in Fig. \ref{F3}. If the number of variables per node is $d$, then the matrix exponential of the Hamiltonian ${{e}^{\text{i}\Delta t\mathbf{K}}}$ is denoted as $\mathbf{G}_{d}^{\left( m \right)}\left( \Delta t\mathbf{K} \right)$. The quantum circuit designed in this study involves two assumptions: 1) the matrix $\mathbf{K}$ can be encoded in a quantum computer; 2) the quantum multiplexer can be implemented in hardware with a complexity of $\text{poly}\left( n \right)$.
		
		\subsection{One-Dimensional Voxel Grid}
		\label{S4S1}
\begin{figure}[!h]
			\centering
			\includegraphics{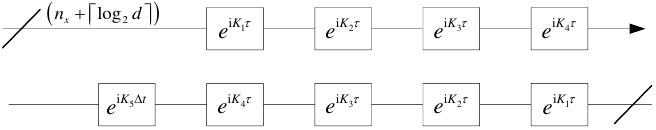}
			\caption{The quantum circuit for simulating ${{e}^{\text{i}\Delta t\mathbf{K}}}$.}
			\label{F6}
		\end{figure}

		First, consider the QHS in the case of the 1-dimensional voxel grids. Place ${{N}_{x}}={{2}^{{{n}_{x}}}}$ points on the solution domain and discretize the domain into ${{N}_{x}}-1$ grids, and the resulting matrix is as shown in Eq. (\ref{E8}). Decompose the matrix via the proposed KCQ decomposition method to obtain 5 matrices ${{\mathbf{K}}_{k}}$ that are easy to simulate. To construct the quantum circuit for simulating ${{e}^{\text{i}\Delta t\mathbf{K}}}$, according to Lemma 1, if third-order accuracy is needed to be achieved, we have,
		\begin{equation}
			\begin{gathered}			
			{{e}^{\text{i}\mathbf{K}\Delta t}}={{e}^{\text{i}{{\mathbf{K}}_{1}}\tau }}{{e}^{\text{i}{{\mathbf{K}}_{2}}\tau }}{{e}^{\text{i}{{\mathbf{K}}_{3}}\tau }}{{e}^{\text{i}{{\mathbf{K}}_{4}}\tau }}{{e}^{\text{i}{{\mathbf{K}}_{5}}\Delta t}}{{e}^{\text{i}{{\mathbf{K}}_{4}}\tau }}{{e}^{\text{i}{{\mathbf{K}}_{3}}\tau }}{{e}^{\text{i}{{\mathbf{K}}_{2}}\tau }}{{e}^{\text{i}{{\mathbf{K}}_{1}}\tau }}+O\left( \Delta {{t}^{3}} \right),\\
			\tau=\frac{\Delta t}{2}.
			\end{gathered}
			\label{E21}
		\end{equation}
		The constructed quantum circuit is shown in Fig. \ref{F6}.
		
		Therefore, the key to simulating ${{e}^{\text{i}\Delta t\mathbf{K}}}$ using quantum computing lies in constructing the corresponding quantum circuit. Next, we will discuss separately according to the different numbers of variables.
		
		\subsubsection{Single-Variable Problem}
		\label{S4S1S1}
		First, consider the single-variable problem, i.e., $d=1$. In this situation, each ${{e}^{\text{i}\Delta t{{\mathbf{K}}_{k}}}}$ is shown in the following equations respectively,
		\begin{equation}
			\begin{aligned}
				& {{e}^{\text{i}\Delta t{{\mathbf{K}}_{1}}}}=\underset{n=0}{\overset{{{2}^{{{n}_{x}}}}-1}{\mathop{\oplus }}}\,{{e}^{\text{i}\Delta t{{k}_{nn}}}}=\underset{n=0}{\overset{{{2}^{{{n}_{x}}-1}}-1}{\mathop{\oplus }}}\,\left[ \begin{matrix}
				{{e}^{\text{i}\Delta t{{k}_{2n,2n}}}} & 0  \\
				0 & {{e}^{\text{i}\Delta t{{k}_{2n+1,2n+1}}}}  \\
				\end{matrix} \right] \\ 
				& =\left[ \underset{n=0}{\overset{{{2}^{{{n}_{x}}-1}}-1}{\mathop{\oplus }}}\,\mathbf{T}\left( \Delta t{{k}_{2n+1,2n+1}} \right) \right]\left[ {{\mathbf{I}}_{{{2}^{{{n}_{x}}-1}}}}\otimes \mathbf{X} \right] \\ 
				& \times \left[ \underset{n=0}{\overset{{{2}^{{{n}_{x}}-1}}-1}{\mathop{\oplus }}}\,\mathbf{T}\left( \Delta t{{k}_{2n,2n}} \right) \right]\left[ {{\mathbf{I}}_{{{2}^{{{n}_{x}}-1}}}}\otimes \mathbf{X} \right]
			\end{aligned},
			\label{E22}
		\end{equation}
		
		\begin{equation}
			\begin{aligned}
				{{e}^{\text{i}\Delta t{{\mathbf{K}}_{2}}}}&=\exp \left[ \text{i}\Delta t\underset{n=0}{\overset{{{2}^{{{n}_{x}}-1}}-1}{\mathop{\oplus }}}\,\left[ \begin{matrix}
					0 & {{c}_{2n}}  \\
					{{c}_{2n}} & 0  \\
				\end{matrix} \right] \right]\\
				& =\exp \left[ \text{i}\Delta t\underset{n=0}{\overset{{{2}^{{{n}_{x}}-1}}-1}{\mathop{\oplus }}}\,\left[ \begin{matrix}
					0 & {{k}_{2n,2n+1}}  \\
					{{k}_{2n,2n+1}} & 0  \\
				\end{matrix} \right] \right] \\ 
				& =\left( {{\mathbf{I}}_{{{2}^{{{n}_{x}}-1}}}}\otimes \mathbf{H} \right)\left[ \underset{n=0}{\overset{{{2}^{{{n}_{x}}-1}}-1}{\mathop{\oplus }}}\,\mathbf{T}\left( -\Delta t{{k}_{2n,2n+1}} \right) \right]\\
				& \times \left( {{\mathbf{I}}_{{{2}^{{{n}_{x}}-1}}}}\otimes \mathbf{X} \right)\underset{n=0}{\overset{{{2}^{{{n}_{x}}-1}}-1}{\mathop{\oplus }}}\,\mathbf{T}\left( \Delta t{{k}_{2n,2n+1}} \right)\left[ {{\mathbf{I}}_{{{2}^{{{n}_{x}}-1}}}}\otimes \left( \mathbf{XH} \right) \right]
			\end{aligned},
			\label{E23}
		\end{equation}
		
		\begin{equation}
			\begin{aligned}
				{{e}^{\text{i}\Delta t{{\mathbf{K}}_{4}}}}&=\mathbf{S}_{{{N}_{x}}}^{\text{H}}\left( {{\mathbf{I}}_{{{2}^{{{n}_{x}}-1}}}}\otimes \mathbf{H} \right)\underset{n=0}{\overset{{{2}^{{{n}_{x}}-1}}-1}{\mathop{\oplus }}}\,\mathbf{T}\left( -\Delta t{{k}_{2n+1,2n+2}} \right)\left( {{\mathbf{I}}_{{{2}^{{{n}_{x}}-1}}}}\otimes \mathbf{X} \right) \\ 
				& \times \underset{n=0}{\overset{{{2}^{{{n}_{x}}-1}}-1}{\mathop{\oplus }}}\,\mathbf{T}\left( \Delta t{{k}_{2n+1,2n+2}} \right)\left[ {{\mathbf{I}}_{{{2}^{{{n}_{x}}-1}}}}\otimes \left( \mathbf{XH} \right) \right]{{\mathbf{S}}_{{{N}_{x}}}} 
			\end{aligned},
			\label{E24}
		\end{equation}
		
		\begin{equation}
			{{e}^{\text{i}\Delta t{{\mathbf{K}}_{3}}}}={{e}^{\text{i}\Delta t{{\mathbf{K}}_{5}}}}={{\mathbf{I}}_{{{N}_{x}}}}.
			\label{E25}
		\end{equation}
		Substituting Eqs. (\ref{E22})-(\ref{E25}) into Eq. (\ref{E21}), it can be obtained
		\begin{equation}
			\begin{aligned}
				{{e}^{\text{i}\mathbf{K}\Delta t}}&=\left[ \underset{n=0}{\overset{{{2}^{{{n}_{x}}-1}}-1}{\mathop{\oplus }}}\,\mathbf{T}\left( \tau {{k}_{2n+1,2n+1}} \right) \right]\left( {{\mathbf{I}}_{{{2}^{{{n}_{x}}-1}}}}\otimes \mathbf{X} \right)\left[ \underset{n=0}{\overset{{{2}^{{{n}_{x}}-1}}-1}{\mathop{\oplus }}}\,\mathbf{T}\left( \tau {{k}_{2n,2n}} \right) \right] \\ 
				& \times \left[ {{\mathbf{I}}_{{{2}^{{{n}_{x}}-1}}}}\otimes \left( \mathbf{XH} \right) \right] \left[ \underset{n=0}{\overset{{{2}^{{{n}_{x}}-1}}-1}{\mathop{\oplus }}}\,\mathbf{T}\left( -\tau {{k}_{2n,2n+1}} \right) \right]\left( {{\mathbf{I}}_{{{2}^{{{n}_{x}}-1}}}}\otimes \mathbf{X} \right) \\ 
				& \times \left[ \underset{n=0}{\overset{{{2}^{{{n}_{x}}-1}}-1}{\mathop{\oplus }}}\,\mathbf{T}\left( \tau {{k}_{2n,2n+1}} \right) \right]\left[ {{\mathbf{I}}_{{{2}^{{{n}_{x}}-1}}}}\otimes \left( \mathbf{XH} \right) \right] \mathbf{S}_{{{N}_{x}}}^{\text{H}}\left( {{\mathbf{I}}_{{{2}^{{{n}_{x}}-1}}}}\otimes \mathbf{H} \right) \\ 
				& \times \left[ \underset{n=0}{\overset{{{2}^{{{n}_{x}}-1}}-1}{\mathop{\oplus }}}\,\mathbf{T}\left( -\Delta t{{k}_{2n+1,2n+2}} \right) \right]\left( {{\mathbf{I}}_{{{2}^{{{n}_{x}}-1}}}}\otimes \mathbf{X} \right) \\ 
				& \times \left[ \underset{n=0}{\overset{{{2}^{{{n}_{x}}-1}}-1}{\mathop{\oplus }}}\,\mathbf{T}\left( \Delta t{{k}_{2n+1,2n+2}} \right) \right] \left[ {{\mathbf{I}}_{{{2}^{{{n}_{x}}-1}}}}\otimes \left( \mathbf{XH} \right) \right]{{\mathbf{S}}_{{{N}_{x}}}} \left( {{\mathbf{I}}_{{{2}^{{{n}_{x}}-1}}}}\otimes \mathbf{H} \right) \\ 
				& \times \left[ \underset{n=0}{\overset{{{2}^{{{n}_{x}}-1}}-1}{\mathop{\oplus }}}\,\mathbf{T}\left( -\tau {{k}_{2n,2n+1}} \right) \right] \left( {{\mathbf{I}}_{{{2}^{{{n}_{x}}-1}}}}\otimes \mathbf{X} \right)\left[ \underset{n=0}{\overset{{{2}^{{{n}_{x}}-1}}-1}{\mathop{\oplus }}}\,\mathbf{T}\left( \tau {{k}_{2n,2n+1}} \right) \right] \\
				& \times \left[ {{\mathbf{I}}_{{{2}^{{{n}_{x}}-1}}}}\otimes \left( \mathbf{XH} \right) \right] \left[ \underset{n=0}{\overset{{{2}^{{{n}_{x}}-1}}-1}{\mathop{\oplus }}}\,\mathbf{T}\left( \tau {{k}_{2n+1,2n+1}} \right) \right]\left[ {{\mathbf{I}}_{{{2}^{{{n}_{x}}-1}}}}\otimes \mathbf{X} \right] \\ 
				& \times \left[ \underset{n=0}{\overset{{{2}^{{{n}_{x}}-1}}-1}{\mathop{\oplus }}}\,\mathbf{T}\left( \tau {{k}_{2n,2n}} \right) \right] \left[ {{\mathbf{I}}_{{{2}^{{{n}_{x}}-1}}}}\otimes \mathbf{X} \right]+O\left( \Delta {{t}^{3}} \right) 
			\end{aligned}.
			\label{E26}
		\end{equation}
		\begin{figure}[!t]
			\centering
			\includegraphics[width=1\textwidth]{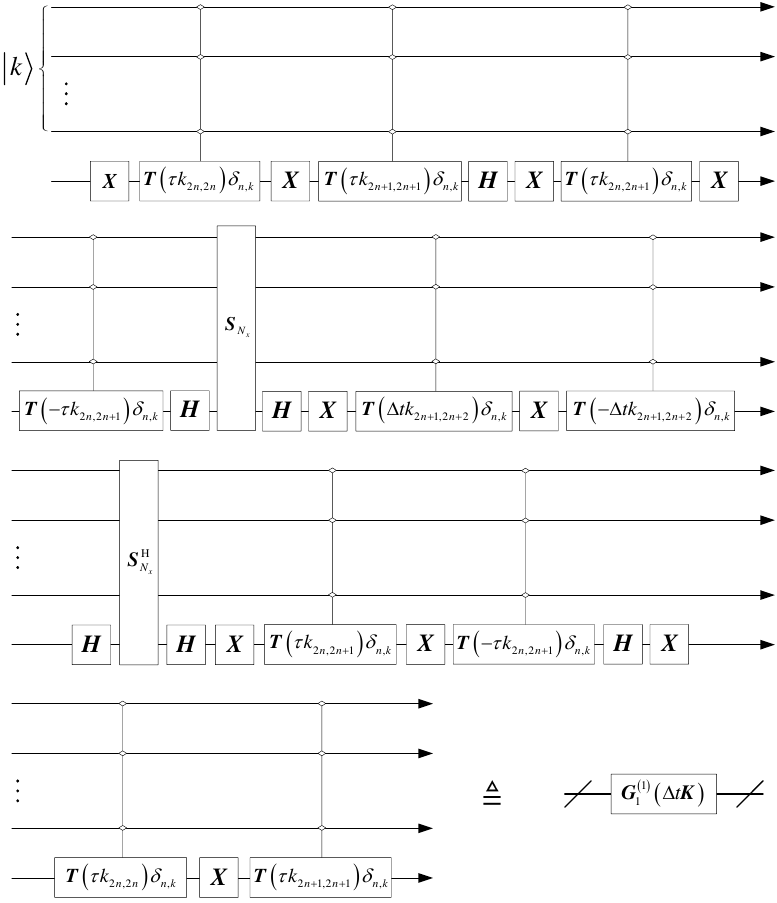}
			\caption{The quantum circuit for simulating $\mathbf{G}_{1}^{\left( 1 \right)}\left( \Delta t\mathbf{K} \right)$.}
			\label{F7}
		\end{figure}
		According to the above equation, the quantum circuit can be drawn as shown in Fig. \ref{F7}. The quantum circuit of the cyclic matrix ${{\mathbf{S}}_{{{N}_{x}}}}$ is shown in \ref{APPA}. By using the KCQ decomposition and Lemma 1, it is easy to construct results with higher precision. The steps are similar, and they will not be elaborated here. 
		
		As can be seen from Figs. \ref{F6}-\ref{F7}, when using the KCQ decomposition method to calculate ${{e}^{\text{i}\Delta t\mathbf{K}}}=\mathbf{G}_{1}^{\left( 1 \right)}\left( \Delta t\mathbf{K} \right)+O\left( \Delta {{t}^{3}} \right)$, the two most resource-consuming parts are ${{\mathbf{S}}_{{{N}_{x}}}}$ and the quantum multiplexer. The quantum circuit of the cyclic matrix ${{\mathbf{S}}_{{{N}_{x}}}}$ is shown in \ref{APPA}. The calculation of ${{\mathbf{S}}_{{{N}_{x}}}}$ involves the QFT whose computational complexity is $O\left( n_{x}^{2} \right)$. If the computational complexity of the quantum multiplexer $\underset{n=0}{\overset{{{2}^{{{n}_{x}}-1}}-1}{\mathop{\oplus }}}\,\mathbf{T}\left( \tau {{k}_{2n,2n}} \right)$ is $\text{poly}\left( {{n}_{x}} \right)$, then the quantum circuit of the QSH of the one-dimensional single-variable problem constructed in Fig. \ref{F7} can be effectively implemented with the complexity of $\text{poly}\left( {{n}_{x}} \right)$. Considering that some 2-qubit quantum multiplexers, as shown in Fig. \ref{F1} (e), have already been implemented on current hardware, and the research on multi-qubit control gates is gradually deepening, it can be anticipated that the implementation of quantum multiplexers on quantum computers is promising in the future.
		
		\subsubsection{Multiple-Variable Problem}
		\label{S4S1S2}
		Consider the multiple-variable problem, so $d=2$ or 4. In this situation, each ${{e}^{\text{i}\Delta t{{\mathbf{K}}_{k}}}}$ is shown in the following equations respectively,
		\begin{equation}
			\begin{aligned}
				& {{e}^{\text{i}\Delta t{{\mathbf{K}}_{1}}}}=\underset{n=0}{\overset{{{2}^{{{n}_{x}}}}-1}{\mathop{\oplus }}}\,{{\mathbf{W}}_{d}}\left( \Delta t{{\mathbf{k}}_{nn}} \right) \\ 
				& {{e}^{\text{i}\Delta t{{\mathbf{K}}_{2}}}}={{\mathbf{L}}_{2,{{N}_{x}},d}}\left[ \underset{n=0}{\overset{{{2}^{{{n}_{x}}}}-1}{\mathop{\oplus }}}\,{{\mathbf{W}}_{d}}\left( \Delta t{{\left( -1 \right)}^{n}}{{\mathbf{c}}_{2\left\lfloor 0.5n \right\rfloor }} \right) \right]\mathbf{L}_{2,{{N}_{x}},d}^{\text{H}} \\ 
				& {{e}^{\text{i}\Delta t{{\mathbf{K}}_{3}}}}={{\mathbf{L}}_{3,{{N}_{x}},d}}\left[ \underset{n=0}{\overset{{{2}^{{{n}_{x}}}}-1}{\mathop{\oplus }}}\,{{\mathbf{W}}_{d}}\left( \Delta t{{\left( -1 \right)}^{n}}{{\mathbf{q}}_{2\left\lfloor 0.5n \right\rfloor }} \right) \right]\mathbf{L}_{3,{{N}_{x}},d}^{\text{H}} \\ 
				& {{e}^{\text{i}\Delta t{{\mathbf{K}}_{4}}}}={{\mathbf{L}}_{4,{{N}_{x}},d}}\left[ \underset{n=0}{\overset{{{2}^{{{n}_{x}}}}-1}{\mathop{\oplus }}}\,{{\mathbf{W}}_{d}}\left( \Delta t{{\left( -1 \right)}^{n}}{{\mathbf{c}}_{2\left\lfloor 0.5n \right\rfloor +1}} \right) \right]\mathbf{L}_{4,{{N}_{x}},d}^{\text{H}} \\ 
				& {{e}^{\text{i}\Delta t{{\mathbf{K}}_{5}}}}={{\mathbf{L}}_{5,{{N}_{x}},d}}\left[ \underset{n=0}{\overset{{{2}^{{{n}_{x}}}}-1}{\mathop{\oplus }}}\,{{\mathbf{W}}_{d}}\left( \Delta t{{\left( -1 \right)}^{n}}{{\mathbf{q}}_{2\left\lfloor 0.5n \right\rfloor +1}} \right) \right]\mathbf{L}_{5,{{N}_{x}},d}^{\text{H}} \\ 
			\end{aligned}
			\label{E27}
		\end{equation}
		Substituting Eqs. (\ref{E27}) into Eq. (\ref{E21}), it can be obtained,
		\begin{equation}
			\begin{aligned}
				 {{e}^{\text{i}\mathbf{K}\Delta t}}& =\mathbf{G}_{d}^{\left( 1 \right)}\left( \Delta t\mathbf{K} \right)+O\left( \Delta {{t}^{3}} \right) \\ 
				& =\left[ \underset{n=0}{\overset{{{2}^{{{n}_{x}}}}-1}{\mathop{\oplus }}}\,{{\mathbf{W}}_{d}}\left( \tau {{\mathbf{k}}_{nn}} \right) \right]{{\mathbf{L}}_{2,{{N}_{x}},d}}\left[ \underset{n=0}{\overset{{{2}^{{{n}_{x}}}}-1}{\mathop{\oplus }}}\,{{\mathbf{W}}_{d}}\left( \tau {{\left( -1 \right)}^{n}}{{\mathbf{c}}_{2\left\lfloor 0.5n \right\rfloor }} \right) \right]\mathbf{L}_{2,{{N}_{x}},d}^{\text{H}} \\ 
				& \times {{\mathbf{L}}_{3,{{N}_{x}},d}}\left[ \underset{n=0}{\overset{{{2}^{{{n}_{x}}}}-1}{\mathop{\oplus }}}\,{{\mathbf{W}}_{d}}\left( \tau {{\left( -1 \right)}^{n}}{{\mathbf{q}}_{2\left\lfloor 0.5n \right\rfloor }} \right) \right]\mathbf{L}_{3,{{N}_{x}},d}^{\text{H}}\\
				& \times {{\mathbf{L}}_{4,{{N}_{x}},d}}\left[ \underset{n=0}{\overset{{{2}^{{{n}_{x}}}}-1}{\mathop{\oplus }}}\,{{\mathbf{W}}_{d}}\left( \tau {{\left( -1 \right)}^{n}}{{\mathbf{c}}_{2\left\lfloor 0.5n \right\rfloor +1}} \right) \right]\mathbf{L}_{4,{{N}_{x}},d}^{\text{H}} \\ 
				& \times {{\mathbf{L}}_{5,{{N}_{x}},d}}\left[ \underset{n=0}{\overset{{{2}^{{{n}_{x}}}}-1}{\mathop{\oplus }}}\,{{\mathbf{W}}_{d}}\left( \Delta t{{\left( -1 \right)}^{n}}{{\mathbf{q}}_{2\left\lfloor 0.5n \right\rfloor +1}} \right) \right]\mathbf{L}_{5,{{N}_{x}},d}^{\text{H}} \\ 
				& \times {{\mathbf{L}}_{4,{{N}_{x}},d}}\left[ \underset{n=0}{\overset{{{2}^{{{n}_{x}}}}-1}{\mathop{\oplus }}}\,{{\mathbf{W}}_{d}}\left( \tau {{\left( -1 \right)}^{n}}{{\mathbf{c}}_{2\left\lfloor 0.5n \right\rfloor +1}} \right) \right]\mathbf{L}_{4,{{N}_{x}},d}^{\text{H}}\\
				& \times {{\mathbf{L}}_{3,{{N}_{x}},d}}\left[ \underset{n=0}{\overset{{{2}^{{{n}_{x}}}}-1}{\mathop{\oplus }}}\,{{\mathbf{W}}_{d}}\left( \tau {{\left( -1 \right)}^{n}}{{\mathbf{q}}_{2\left\lfloor 0.5n \right\rfloor }} \right) \right]\mathbf{L}_{3,{{N}_{x}},d}^{\text{H}} \\ 
				& \times {{\mathbf{L}}_{2,{{N}_{x}},d}}\left[ \underset{n=0}{\overset{{{2}^{{{n}_{x}}}}-1}{\mathop{\oplus }}}\,{{\mathbf{W}}_{d}}\left( \tau {{\left( -1 \right)}^{n}}{{\mathbf{c}}_{2\left\lfloor 0.5n \right\rfloor }} \right) \right]\mathbf{L}_{2,{{N}_{x}},d}^{\text{H}}\\
				& \times \left[ \underset{n=0}{\overset{{{2}^{{{n}_{x}}}}-1}{\mathop{\oplus }}}\,{{\mathbf{W}}_{d}}\left( \tau {{\mathbf{k}}_{nn}} \right) \right]+O\left( \Delta {{t}^{3}} \right),\ \ \ \tau =\frac{\Delta t}{2} 
			\end{aligned}.
			\label{E28}
		\end{equation}
		
\begin{figure}[!h]
			\centering
			\includegraphics[width=1\textwidth]{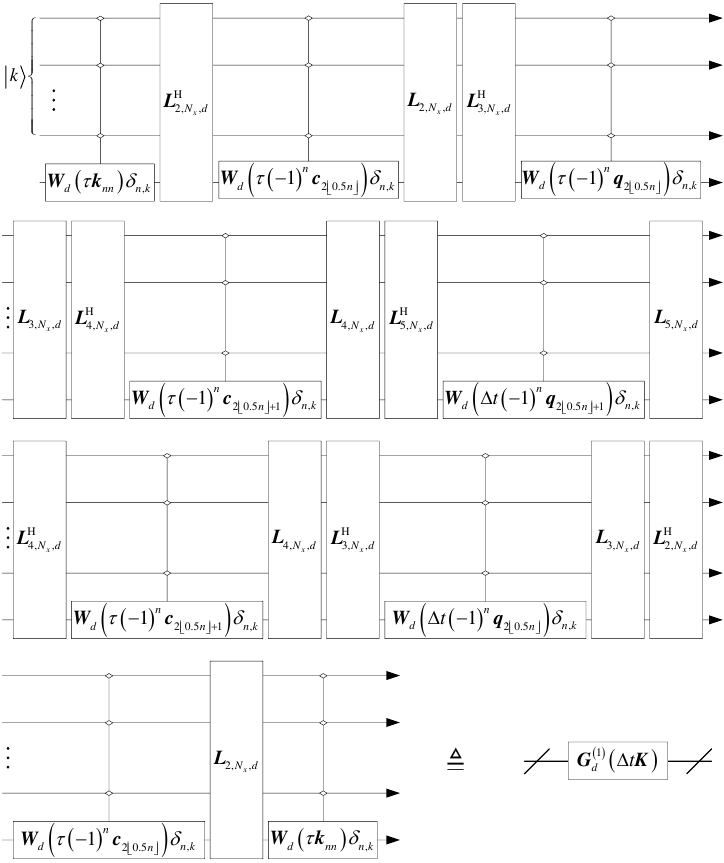}
			\caption{The quantum circuit for simulating $\mathbf{G}_{d}^{\left( 1 \right)}\left( \Delta t\mathbf{K} \right)$.}
			\label{F8}
		\end{figure}

		Based on the above equation, the quantum circuit can be drawn as shown in Fig. \ref{F8}. It can be seen that when using the KCQ decomposition method to calculate ${{e}^{\text{i}\mathbf{K}\Delta t}}=\mathbf{G}_{d}^{\left( 1 \right)}\left( \Delta t\mathbf{K} \right)+O\left( \Delta {{t}^{3}} \right)$ under the assumptions that the quantum multiplexer can be implemented in hardware, the computational complexity is $\text{poly}\left( {{n}_{x}}+d \right)$.
		
		\subsection{Two-Dimensional Voxel Grids}
		\label{S4S2}
		By utilizing the tridiagonal fractal property of the MVG and the KCQ decomposition method, the QSH of two-dimensional or three-dimensional voxel grids can be easily achieved. Taking a two-dimensional voxel grid as an example, the KCQ decomposition form is shown in Eq. (\ref{E18}). Establish VBQC for each sub-matrix respectively. First, let's study ${{\mathbf{K}}_{1}}=\underset{n=0}{\overset{{{2}^{{{n}_{x}}}}-1}{\mathop{\oplus }}}\,{{\mathbf{k}}_{nn}}\left( \mathbf{K} \right)$, whose matrix exponential can be written as,
		\begin{equation}
			{{e}^{\text{i}\Delta t{{\mathbf{K}}_{1}}}}=\underset{n=0}{\overset{{{2}^{{{n}_{x}}}}-1}{\mathop{\oplus }}}\,{{e}^{\text{i}\Delta t{{\mathbf{k}}_{nn}}\left( \mathbf{K} \right)}}
			\label{E29}.
		\end{equation}
		Note that ${{\mathbf{k}}_{nn}}\left( \mathbf{K} \right)$ is a matrix generated by 1-dimensional voxel grids and also a tridiagonal block matrix. Therefore, the matrix exponential of ${{\mathbf{k}}_{nn}}\left( \mathbf{K} \right)$ can be calculated using the quantum circuit established in subsection \ref{S4S1}, which is $\mathbf{G}_{d}^{\left( 1 \right)}\left( \Delta t{{\mathbf{k}}_{nn}}\left( \mathbf{K} \right) \right)$. So it can be obtained,
		\begin{equation}
			{{e}^{\text{i}\Delta t{{\mathbf{K}}_{1}}}}=\underset{n=0}{\overset{{{2}^{{{n}_{x}}}}-1}{\mathop{\oplus }}}\,\mathbf{G}_{d}^{\left( 1 \right)}\left( \Delta t{{\mathbf{k}}_{nn}}\left( \mathbf{K} \right) \right)
			\label{E30}.
		\end{equation}
		Furthermore, note that the structural forms of both ${{\mathbf{c}}_{n}}\left( \mathbf{K} \right)$ and ${{\mathbf{q}}_{n}}\left( \mathbf{K} \right)$ are the same as that of the matrices generated by 1-dimensional voxel grids. Thus, both ${{\mathbf{c}}_{n}}\left( \mathbf{K} \right)$ and ${{\mathbf{q}}_{n}}\left( \mathbf{K} \right)$ are also tridiagonal block Hermitian matrices. Therefore, their matrix exponentials can also be calculated using the quantum circuit established in subsection \ref{S4S1}. Then, we have
		\begin{equation}
			\begin{aligned}
				& {{e}^{\text{i}\Delta t{{\mathbf{K}}_{2}}}}={{\mathbf{L}}_{2,{{N}_{x}},{{N}_{y}}d}}\left[ \underset{n=0}{\overset{{{2}^{{{n}_{x}}}}-1}{\mathop{\oplus }}}\,\mathbf{G}_{d}^{\left( 1 \right)}\left( \Delta t{{\left( -1 \right)}^{n}}{{\mathbf{c}}_{2\left\lfloor 0.5n \right\rfloor }}\left( \mathbf{K} \right) \right) \right]\mathbf{L}_{2,{{N}_{x}},{{N}_{y}}d}^{\text{H}} \\ 
				& {{e}^{\text{i}\Delta t{{\mathbf{K}}_{3}}}}={{\mathbf{L}}_{3,{{N}_{x}},{{N}_{y}}d}}\left[ \underset{n=0}{\overset{{{2}^{{{n}_{x}}}}-1}{\mathop{\oplus }}}\,\mathbf{G}_{d}^{\left( 1 \right)}\left( \Delta t{{\left( -1 \right)}^{n}}{{\mathbf{q}}_{2\left\lfloor 0.5n \right\rfloor }}\left( \mathbf{K} \right) \right) \right]\mathbf{L}_{3,{{N}_{x}},{{N}_{y}}d}^{\text{H}} \\ 
				& {{e}^{\text{i}\Delta t{{\mathbf{K}}_{4}}}}={{\mathbf{L}}_{4,{{N}_{x}},{{N}_{y}}d}}\left[ \underset{n=0}{\overset{{{2}^{{{n}_{x}}}}-1}{\mathop{\oplus }}}\,\mathbf{G}_{d}^{\left( 1 \right)}\left( \Delta t{{\left( -1 \right)}^{n}}{{\mathbf{c}}_{2\left\lfloor 0.5n \right\rfloor +1}}\left( \mathbf{K} \right) \right) \right]\mathbf{L}_{4,{{N}_{x}},{{N}_{y}}d}^{\text{H}} \\ 
				& {{e}^{\text{i}\Delta t{{\mathbf{K}}_{5}}}}={{\mathbf{L}}_{5,{{N}_{x}},{{N}_{y}}d}}\left[ \underset{n=0}{\overset{{{2}^{{{n}_{x}}}}-1}{\mathop{\oplus }}}\,\mathbf{G}_{d}^{\left( 1 \right)}\left( \Delta t{{\left( -1 \right)}^{n}}{{\mathbf{q}}_{2\left\lfloor 0.5n \right\rfloor +1}}\left( \mathbf{K} \right) \right) \right]\mathbf{L}_{5,{{N}_{x}},{{N}_{y}}d}^{\text{H}} \\ 
			\end{aligned}
			\label{E31}.
		\end{equation}
		
		Substitute Eq. (\ref{E31}) into Eq. (\ref{E21}), and we can obtain
		\begin{equation}
			\begin{aligned}
				{{e}^{\text{i}\mathbf{K}\Delta t}}&=\mathbf{G}_{d}^{\left( 2 \right)}\left( \Delta t\mathbf{K} \right)+O\left( \Delta {{t}^{3}} \right) \\ 
				& =\left[ \underset{n=0}{\overset{{{2}^{{{n}_{x}}}}-1}{\mathop{\oplus }}}\,\mathbf{G}_{d}^{\left( 1 \right)}\left( \tau {{\mathbf{k}}_{nn}}\left( \mathbf{K} \right) \right) \right] \\ 
				& \times {{\mathbf{L}}_{2,{{N}_{x}},{{N}_{y}}d}} \left[ \underset{n=0}{\overset{{{2}^{{{n}_{x}}}}-1}{\mathop{\oplus }}}\,\mathbf{G}_{d}^{\left( 1 \right)}\left( \tau {{\left( -1 \right)}^{n}}{{\mathbf{c}}_{2\left\lfloor 0.5n \right\rfloor }}\left( \mathbf{K} \right) \right) \right] \mathbf{L}_{2,{{N}_{x}},{{N}_{y}}d}^{\text{H}} \\
				& \times {{\mathbf{L}}_{3,{{N}_{x}},{{N}_{y}}d}}\left[ \underset{n=0}{\overset{{{2}^{{{n}_{x}}}}-1}{\mathop{\oplus }}}\,\mathbf{G}_{d}^{\left( 1 \right)}\left( \tau {{\left( -1 \right)}^{n}}{{\mathbf{q}}_{2\left\lfloor 0.5n \right\rfloor }}\left( \mathbf{K} \right) \right) \right]\mathbf{L}_{3,{{N}_{x}},{{N}_{y}}d}^{\text{H}}\\
				& \times {{\mathbf{L}}_{4,{{N}_{x}},{{N}_{y}}d}}\left[ \underset{n=0}{\overset{{{2}^{{{n}_{x}}}}-1}{\mathop{\oplus }}}\,\mathbf{G}_{d}^{\left( 1 \right)}\left( \tau {{\left( -1 \right)}^{n}}{{\mathbf{c}}_{2\left\lfloor 0.5n \right\rfloor +1}}\left( \mathbf{K} \right) \right) \right] \mathbf{L}_{4,{{N}_{x}},{{N}_{y}}d}^{\text{H}} \\ 
				& \times {{\mathbf{L}}_{5,{{N}_{x}},{{N}_{y}}d}}\left[ \underset{n=0}{\overset{{{2}^{{{n}_{x}}}}-1}{\mathop{\oplus }}}\,\mathbf{G}_{d}^{\left( 1 \right)}\left( \Delta t{{\left( -1 \right)}^{n}}{{\mathbf{q}}_{2\left\lfloor 0.5n \right\rfloor +1}}\left( \mathbf{K} \right) \right) \right]\mathbf{L}_{5,{{N}_{x}},{{N}_{y}}d}^{\text{H}} \\ 
				& \times {{\mathbf{L}}_{4,{{N}_{x}},{{N}_{y}}d}} \left[ \underset{n=0}{\overset{{{2}^{{{n}_{x}}}}-1}{\mathop{\oplus }}}\,\mathbf{G}_{d}^{\left( 1 \right)}\left( \tau {{\left( -1 \right)}^{n}}{{\mathbf{c}}_{2\left\lfloor 0.5n \right\rfloor +1}}\left( \mathbf{K} \right) \right) \right]\mathbf{L}_{4,{{N}_{x}},{{N}_{y}}d}^{\text{H}} \\
				& \times {{\mathbf{L}}_{3,{{N}_{x}},{{N}_{y}}d}}\left[ \underset{n=0}{\overset{{{2}^{{{n}_{x}}}}-1}{\mathop{\oplus }}}\,\mathbf{G}_{d}^{\left( 1 \right)}\left( \tau {{\left( -1 \right)}^{n}}{{\mathbf{q}}_{2\left\lfloor 0.5n \right\rfloor }}\left( \mathbf{K} \right) \right) \right]\mathbf{L}_{3,{{N}_{x}},{{N}_{y}}d}^{\text{H}} \\ 
				& \times {{\mathbf{L}}_{2,{{N}_{x}},{{N}_{y}}d}}\left[ \underset{n=0}{\overset{{{2}^{{{n}_{x}}}}-1}{\mathop{\oplus }}}\,\mathbf{G}_{d}^{\left( 1 \right)}\left( \tau {{\left( -1 \right)}^{n}}{{\mathbf{c}}_{2\left\lfloor 0.5n \right\rfloor }}\left( \mathbf{K} \right) \right) \right]\mathbf{L}_{2,{{N}_{x}},{{N}_{y}}d}^{\text{H}}\\
				& \times \left[ \underset{n=0}{\overset{{{2}^{{{n}_{x}}}}-1}{\mathop{\oplus }}}\,\mathbf{G}_{d}^{\left( 1 \right)}\left( \tau {{\mathbf{k}}_{nn}}\left( \mathbf{K} \right) \right) \right]+O\left( \Delta {{t}^{3}} \right) \\ 
			\end{aligned}
			\label{E32}.
		\end{equation}
		
\begin{figure}[!h]
			\centering
			\includegraphics[width=1\textwidth]{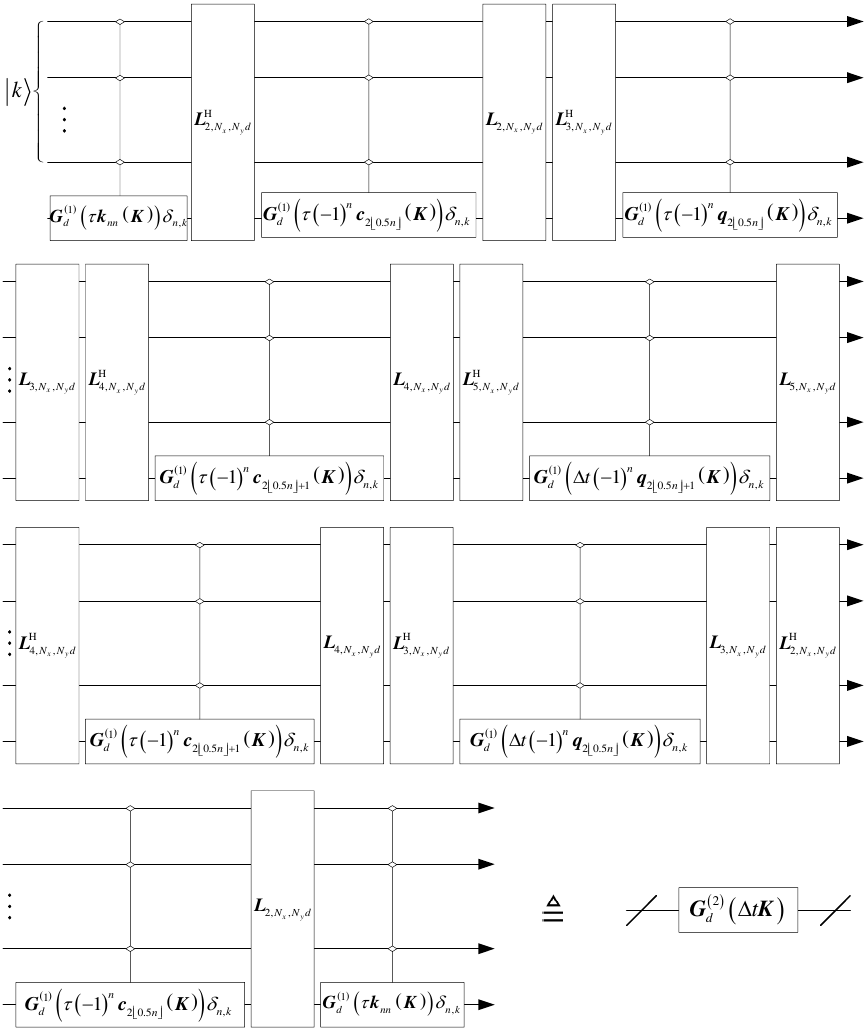}
			\caption{The quantum circuit for simulating $\mathbf{G}_{d}^{\left( 2 \right)}\left( \Delta t\mathbf{K} \right)$.}
			\label{F9}
		\end{figure}

		Based on the above equation, the quantum circuit can be drawn as shown in Fig. \ref{F9}. As can be seen from Fig. \ref{F9}, when using the KCQ decomposition method to calculate ${{e}^{\text{i}\mathbf{K}\Delta t}}=\mathbf{G}_{d}^{\left( 2 \right)}\left( \Delta t\mathbf{K} \right)+O\left( \Delta {{t}^{3}} \right)$, under the conditions assumed in this paper, the computational complexity is $\text{poly}\left( {{n}_{x}}+{{n}_{y}}+d \right)$. By utilizing the tridiagonal fractal property of the MVG and the KCQ decomposition, the QSH of 3-dimensional voxel grids can be easily achieved. The circuit is similar to that in Fig. \ref{F9}, so it will not be elaborated.
		
		\section{Quantum Simulation Examples}
		\label{S5}
		This section verifies the VBQC and the corresponding quantum circuits designed in Section \ref{S4}. Since some circuits in the proposed algorithm cannot yet be implemented on the quantum computer, the constructed quantum circuits are simulated on the classical computer. The considered examples include a 1-dimensional beam, a 2-dimensional plate with holes, and a 3-dimensional inhomogeneous solid problem. For these three common solid mechanics problems, we first obtain the stiffness matrix $\mathbf{K}$ through finite element discretization. Then, we use the proposed VBQC for the QSH in general solid mechanics to compute the matrix exponential ${{e}^{\text{i}\mathbf{K}\Delta t}}$ and compare the results with reference solutions. The reference solutions are calculated by using the precise integration method \citep{R44,R45} commonly used in classical dynamics, and the 1-norm of the difference between the results of the VBQC and the reference solutions is used to measure the error of the proposed algorithm. To further verify the correctness of the proposed VBQC, we use the quantum phase estimation method to compute the natural frequencies and mode shapes of the three problems. The quantum phase estimation method, whose quantum circuit is shown in \ref{APPC}, is a quantum algorithm for solving eigenvalue problems. In the implementation of the examples, we consider the eigenvalue problem with the lumped mass matrix $\mathbf{M}$ in the following form,
		\begin{equation}
			\mathbf{\bar{K}}{{\mathbf{y}}_{i}}=\omega _{i}^{2}{{\mathbf{y}}_{i}},\ \ \ {{\mathbf{x}}_{i}}={{\mathbf{M}}^{-0.5}}{{\mathbf{y}}_{i}},\ \ \ \mathbf{\bar{K}}={{\mathbf{M}}^{-0.5}}\mathbf{K}{{\mathbf{M}}^{-0.5}},
			\label{E33}
		\end{equation}
		where ${{\omega }_{i}}$ and ${{\mathbf{x}}_{i}}$ denote the natural vibration frequency and mode shape, respectively. According to quantum phase estimation, the computation of QSH (i.e., the matrix exponential ${{e}^{\text{i}\mathbf{\bar{K}}\Delta t}}$) is critical. In this section, the proposed VBQC for the QSH is used to compute ${{e}^{\text{i}\mathbf{\bar{K}}\Delta t}}$, and then the quantum phase estimation method is employed to calculate the natural frequencies and mode shapes, which are compared with reference solutions obtained directly from classical finite element calculations.
		
		\subsection{Euler-Bernoulli Beams Fixed Supported at Both Ends}
		\label{S5S1}
		Consider an Euler-Bernoulli beam that has fixed supports at both ends governed by
		\begin{equation}
			\frac{{{\partial }^{4}}w\left( x,\ t \right)}{\partial {{x}^{4}}}+\rho \left( x \right)\frac{{{\partial }^{2}}w\left( x,\ t \right)}{\partial {{t}^{2}}}=0,
			\label{E34}
		\end{equation}
		where
		\begin{equation}
			w\left( 0 \right)=w\left( 1 \right)=\theta \left( 0 \right)=\theta \left( 1 \right)=0,\ \ \ \rho \left( x \right)=\left\{ \begin{aligned}
				& 1,\ \ \ x\in \left[ 0,\ 0.5 \right], \\ 
				& 2,\ \ \ x\in \left[ 0.5,1 \right]. \\ 
			\end{aligned} \right.
			\label{E35}
		\end{equation}
		
		\subsubsection{Accuracy Verification of the QSH}
		\label{S5S1S1}
		\begin{figure}[!h]
			\centering
			\includegraphics{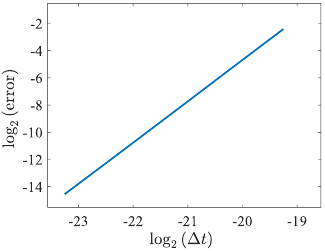}
			\caption{Error analysis of the quantum computation for ${{e}^{\text{i}\mathbf{K}\Delta t}}$.}
			\label{F10}
		\end{figure}
		First, the correctness of the proposed VBQC for the QSH based on the KCQ decomposition method is verified by directly computing the matrix exponential. The ${{2}^{5}}+1$ grids are used for discretization, where each node contains two degrees of freedom (displacement and rotation angle). Therefore, performing the QSH using the proposed method requires a total of 6 qubits. After generating the stiffness matrix $\mathbf{K}$, different time step sizes $\Delta t$ are taken. VBQC is used to compute ${{e}^{\text{i}\mathbf{K}\Delta t}}$, and the results are compared with reference solutions, as shown in Fig. \ref{F10}. It is indicated from Fig. \ref{F10} that as the time step size $\Delta t$ decreases, the error of the quantum computation of ${{e}^{\text{i}\mathbf{K}\Delta t}}$ gradually decreases, with the error order being $O\left( \Delta {{t}^{3.03}} \right)$, which is consistent with the theoretical error order, thereby demonstrating the correctness of the proposed method.
	
		\subsubsection{Natural Vibration Frequency}
		\label{S5S1S2}
\begin{figure}[!h]
			\centering
			\includegraphics{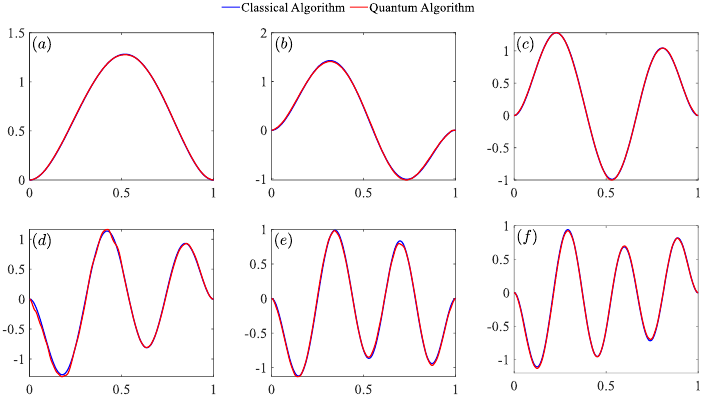}
			\caption{Comparison of mode shapes calculated by classical and quantum algorithms: (a) 1st-order mode shape; (b) 2nd-order mode shape; (c) 3rd-order mode shape; (d)4th-order mode shape; (e) 5th-order mode shape; (f) 6th-order mode shape.}
			\label{F11}
		\end{figure}

		Next, the correctness of the proposed method in analyzing the free vibration of the beam is further verified through the quantum phase estimation. The quantum phase estimation method is used to solve for the first six-order natural vibration frequencies and mode shapes, where QSH uses 6 qubits and the number of auxiliary qubits is 9, so a total of 15 qubits are used in the considered example. The effective computation of ${{e}^{\text{i}\mathbf{K}\Delta t}}$ is the core step, which is solved using the proposed VBQC for the QSH based on the KCQ decomposition method. The comparison between the results of the quantum algorithm and the classical algorithm is presented in Table \ref{T2}, and the mode shapes calculated by different methods are shown in Fig. \ref{F11}.
		\begin{table}[!h]
			\caption{Comparison of the First Six-Order Natural Vibration Frequencies Calculated by Classical and Quantum Algorithms.}
			\label{T2}
			\renewcommand\arraystretch{1.2} 
			\setlength{\tabcolsep}{2.25mm}
			\centering
			\begin{tabular}{ccccccc}
				\hline
				Order & 1 & 2 & 3 & 4 & 5 & 6 \\
				\hline
				Classical Algorithm & 18.1 & 49.5 & 90.6 & 141.2 & 195.7 & 251.9 \\
				Quantum Algorithm & 18.0 & 48.5 & 90.2 & 141.6 & 196.0 & 252.0 \\
				Relative Error & 0.45\% & 1.99\% & 0.45\% & 0.24\% & 0.16\% & 0.04\% \\
				\hline 
			\end{tabular}
		\end{table}
		
		By comparing the natural vibration frequencies calculated by the classical algorithm and the proposed quantum algorithm in Table \ref{T2}, it can be seen that the relative errors are mostly at a low level, with none exceeding 2\% for each order of natural frequencies. Except for the second-order natural frequency, which has a relative error of 1.99\%, the errors for all other orders are less than 0.5\%. Figure \ref{F11} shows that the mode shape curves simulated by the classical and quantum algorithms basically coincide, further indicating the high accuracy of the quantum algorithm in calculating natural vibration frequencies. Table \ref{T2} and Fig. \ref{F11} fully demonstrate that the proposed VBQC can effectively analyze the free vibration of 1-dimensional beams.
		
		\subsection{Four-Sides-Fixed 2-dimensional Perforated Plate}
		\label{S5S2}
\begin{figure}[!h]
			\centering
			\includegraphics{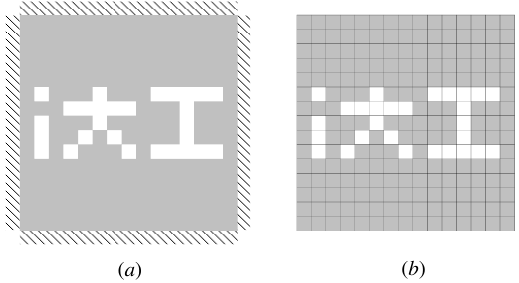}
			\caption{Four-Sides-Fixed 2-dimensional Perforated Plate: (a) Schematic diagram of the model; (b) Voxel grid discretization.}
			\label{F12}
		\end{figure}

		Next, consider a four-sides-fixed 2-dimensional perforated plate, as shown in Fig. \ref{F12} (a). This is a three-variable problem, where the variables are the deflection $w$ and the rotation angles ${{\theta }_{x}}$, ${{\theta }_{y}}$, so a virtual variable is introduced for each node. For ease of calculation, the parameters of the plate are taken in dimensionless form: the side length is 15, the thickness is 1, Young’s modulus is 1, Poisson’s ratio is 0.5, and the density is 1. The Mindlin plate element is used for analysis, and the plate is discretized using the voxel grid, as shown in Fig. \ref{F12} (b). The holes correspond to virtual plate elements with a density of 0 and Young’s modulus of 0.
		
		\subsubsection{Accuracy Verification of Hamiltonian Simulation}
		\label{S5S2S1}
\begin{figure}[!h]
			\centering
			\includegraphics{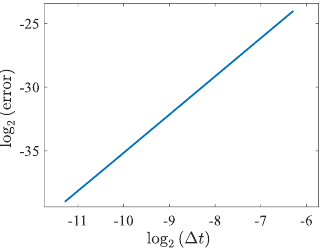}
			\caption{Error analysis of the quantum computation for ${{e}^{\text{i}\mathbf{K}\Delta t}}$.}
			\label{F13}
		\end{figure}

		First, the correctness of the proposed VBQC is verified by directly computing the matrix exponential. According to the voxel grid shown in Fig. \ref{F12} (b), the number of nodes in the considered example is ${{2}^{4}}\times {{2}^{4}}$ and the number of variables is 3. Therefore, performing the QSH using the proposed method requires a total of 10 qubits. After generating the stiffness matrix $\mathbf{K}$ from the grid shown in Fig. \ref{F12} (b), different time step sizes $\Delta t$ are used to compute ${{e}^{\text{i}\mathbf{K}\Delta t}}$ via the proposed quantum algorithm, and the results are compared with reference solutions. The calculation errors of the proposed quantum algorithm under different time step sizes are shown in Fig. \ref{F13}. It can be seen from Fig. \ref{F13} that as the time step size $\Delta t$ decreases, the error of the quantum computation of ${{e}^{\text{i}\mathbf{K}\Delta t}}$ gradually decreases, with an error order of $O\left( \Delta {{t}^{2.99}} \right)$, which is consistent with the theoretical error order, thereby demonstrating the correctness of the proposed method.
		
		\subsubsection{Natural Vibration Frequency}
		\label{S5S2S2}
\begin{figure}[!h]
			\centering
			\includegraphics[width=1\textwidth]{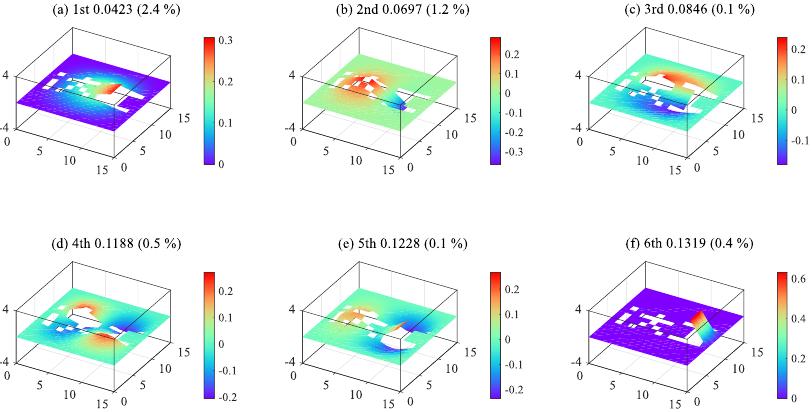}
			\caption{The first six-order mode shapes, natural vibration frequencies, and mode shape errors of the 2-dimensional perforated plate calculated by the quantum algorithm.}
			\label{F14}
		\end{figure}

		\begin{table}[!h]
			\caption{Comparison of the First Six-Order Natural Vibration Frequencies Calculated by Classical and Quantum Algorithms.}
			\label{T3}
			\renewcommand\arraystretch{1.2} 
			\setlength{\tabcolsep}{2.25mm}
			\centering
			\begin{tabular}{ccccccc}
				\hline
				Order & 1 & 2 & 3 & 4 & 5 & 6 \\
				\hline
				Classical Algorithm & 0.0409 & 0.0694 & 0.0843 & 0.1188 & 0.1227 & 0.1315 \\
				Quantum Algorithm & 0.0423 & 0.0697 & 0.0846 & 0.1186 & 0.1228 & 0.1319 \\
				Relative Error & 3.36\% & 0.38\% & 0.34\% & 0.16\% & 0.08\% & 0.24\% \\
				\hline 
			\end{tabular}
		\end{table}
		The correctness of the proposed quantum algorithm for analyzing the free vibration of the 2-dimensional perforated plate is further verified through quantum phase estimation. Natural frequencies and mode shapes are solved using quantum phase estimation, where the computation of ${{e}^{\text{i}\mathbf{K}\Delta t}}$ relies on the proposed VBQC. The QSH uses 10 qubits and the number of auxiliary qubits is 9, so a total of 19 qubits are used in the considered example. The comparison between the results of the quantum algorithm and the classical algorithm is listed in Table \ref{T3}, and the mode shapes calculated by the proposed quantum algorithm are shown in Fig. \ref{F14}, where the results in parentheses denote the relative errors between the mode shapes calculated by the proposed quantum algorithm and the classical algorithm.
		
		By comparing the calculated natural vibration frequencies between the classical and quantum algorithms across different orders in Table \ref{T3}, it can be seen that most relative errors remain at a low level, with none exceeding 4\% for each order of natural frequencies. Figure \ref{F14} shows that the relative errors of the mode shapes for the first six orders of the plate calculated by the quantum algorithm are all less than 3\%, further indicating that the proposed quantum algorithm exhibits high accuracy in computing both the vibration frequencies and mode shapes of the considered plate.
		
		\subsection{3-dimensional Inhomogeneous Solid Problem}
		\label{S5S3}
\begin{figure}[!h]
			\centering
			\includegraphics{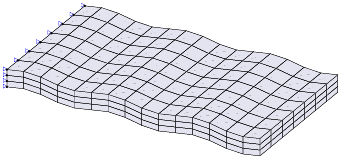}
			\caption{Schematic diagram of the 3-dimensional solid wavy plate with fixed support at the left end.}
			\label{F15}
		\end{figure}

		The third example considers a 3-dimensional wavy plate problem with a fixed support at the left end, as shown in Fig. \ref{F15}. It is a three-variable problem where the variables to be solved are the displacements $u$, $v$, and $w$ in the three coordinate directions, requiring the introduction of a virtual variable for each node. For ease of calculation, the plate parameters are defined in dimensionless form: the plate has a length of 15, width of 7, and thickness of 1. Young’s modulus is given by $100+{{\left( \sin \left( \frac{2\pi }{5}r \right)+\cos \left( \frac{2\pi }{10}r \right) \right)}^{2}}$, where $r=\sqrt{{{x}^{2}}+{{y}^{2}}+{{z}^{2}}}$ and $x$, $y$, $z$ are the initial coordinates of the wavy plate in the three directions. The Poisson’s ratio is 0.3, and the density is 1. The problem is analyzed using 3-dimensional 8-node hexahedral solid elements and discretized with a grid homeomorphic to the voxel grids, as illustrated in Fig. \ref{F15}.
		
		\subsubsection{Accuracy Verification of Hamiltonian Simulation}
		\label{S5S3S1}
		\begin{figure}[!h]
			\centering
			\includegraphics{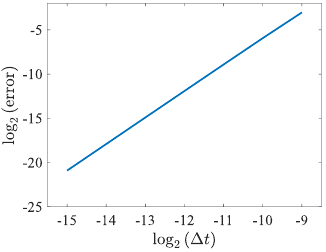}
			\caption{Error analysis of the quantum computation for ${{e}^{\text{i}\mathbf{K}\Delta t}}$.}
			\label{F16}
		\end{figure}
		First, the accuracy of the proposed VBQC for calculating the matrix exponential in 3-dimensional three-variable problems is analyzed. According to the voxel grid shown in Fig. \ref{F15}, the number of nodes in the considered example is ${{2}^{4}}\times {{2}^{3}}\times {{2}^{2}}$ and the number of variables is 3. Therefore, performing the QSH using the proposed method requires a total of 11 qubits. After generating the stiffness matrix $\mathbf{K}$ from the grid shown in Fig. \ref{F15}, different time step sizes $\Delta t$ are used to compute ${{e}^{\text{i}\mathbf{K}\Delta t}}$ via the proposed QSH algorithm based on KCQ decomposition, with results compared to those from classical algorithms. The classical algorithm employs the precise integration method, and errors are calculated using the 1-norm, as shown in Fig. \ref{F16}. It can be observed from Fig. \ref{F16} that as the time step size $\Delta t$ decreases, the error of computing ${{e}^{\text{i}\mathbf{K}\Delta t}}$ using the proposed quantum algorithm gradually decreases, with an error order of $O\left( \Delta {{t}^{2.98}} \right)$, which is consistent with the theoretical error order, thereby demonstrating the correctness of the proposed algorithm.
		
		\subsubsection{Natural Vibration Frequency}
		\label{S5S3S2}
\begin{figure}[!h]
            \centering
            \includegraphics[width=1\textwidth]{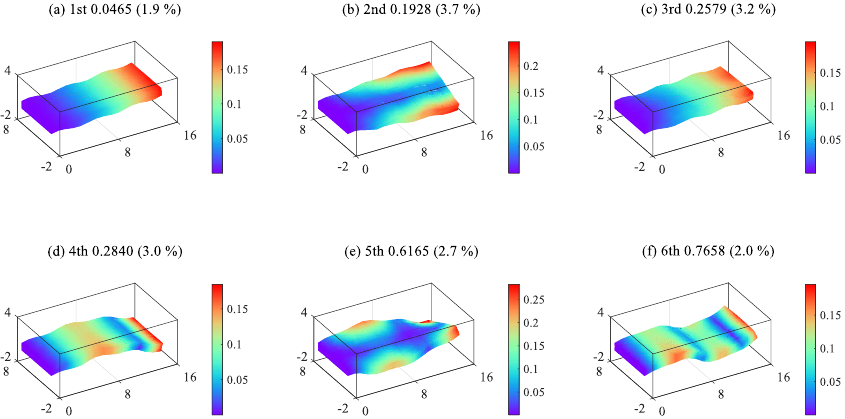}
            \caption{The first six-order mode shapes, natural vibration frequencies, and mode shape errors of the 3-dimensional wavy plate calculated by the quantum algorithm.}
            \label{F17}
        \end{figure}

		Next, the quantum phase estimation method is further used to solve the first six-order natural vibration frequencies and mode shapes of the considered plate. The ${{e}^{\text{i}\mathbf{K}\Delta t}}$ is computed using the proposed VBQC for the QSH based on the KCQ decomposition. The QSH uses 11 qubits and the number of auxiliary qubits is 9, so a total of 20 qubits are used in the considered example. The comparison of the natural vibration frequencies obtained by the quantum algorithm with the classical algorithm is listed in Table \ref{T4}. The mode shapes of each order obtained by the quantum algorithm are shown in Fig. \ref{F17}, where the numbers in parentheses denote the relative errors between the mode shapes obtained by the quantum algorithm and the classical algorithm.
		\begin{table}[!h]
			\caption{Comparison of the First Six-Order Natural Vibration Frequencies Calculated by Classical and Quantum Algorithms.}
			\label{T4}
			\renewcommand\arraystretch{1.2} 
			\setlength{\tabcolsep}{2.25mm}
			\centering
			\begin{tabular}{ccccccc}
				\hline
				Order & 1 & 2 & 3 & 4 & 5 & 6 \\
				\hline
				Classical Algorithm & 0.0495 & 0.1919 & 0.2574 & 0.2846 & 0.6168 & 0.7659 \\
				Quantum Algorithm & 0.0465 & 0.1928 & 0.2579 & 0.2840 & 0.6165 & 0.7658 \\
				Relative Error &6.43\% & 0.46\% & 0.19\% & 0.23\% & 0.05\% & 0.01\% \\
				\hline 
			\end{tabular}
		\end{table}
		
		By comparing the natural vibration frequencies calculated by the classical and quantum algorithms across different orders in Table \ref{T4}, it can be seen that most relative errors remain at a low level, indicating that the proposed quantum algorithm exhibits high accuracy in computing the natural vibration frequencies of the considered problem. Figure \ref{F17} shows that the errors between the first six-order mode shapes of the wavy plate calculated by the quantum algorithm and those by the classical algorithm are all within 4\%, further demonstrating the high accuracy of the proposed quantum algorithm in calculating the mode shapes. The calculation results indicate that the proposed VBQC for solid mechanics problems can effectively simulate natural vibration frequencies of the considered problems, preliminarily verifying the feasibility in solving solid mechanics problems and providing new thinking and practice for the application of quantum computing in the field of engineering calculations.
		
		\section{Conclusion}
		\label{S6}
		This paper conducts in-depth research on the core problem of quantum simulation of Hamiltonians in solid mechanics, specifically the matrix exponential ${{e}^{\text{i}\mathbf{A}t}}$. First, by introducing the concept of virtual nodes, we apply voxel grids to structural modeling and analyze the tridiagonal fractal property of finite element matrices based on voxel grids. On this basis, a KCQ decomposition method is proposed for the MVG by integrating cyclic matrices, matrix direct products, direct sums, and Pauli matrices. The proposed decomposition method can break down the matrix derived from the voxel grid into three sets of fundamental matrices ${{\mathbf{k}}_{n}}$, ${{\mathbf{c}}_{n}}$, and ${{\mathbf{q}}_{n}}$ with dimensions $d=1,\ 2,\ \text{or}\ 4$, thereby reducing the computation of ${{e}^{\text{i}\mathbf{A}t}}$ to calculating the matrix exponentials of these three sets of $d\text{-dimensional}$ matrices. Subsequently, based on the KCQ decomposition and leveraging QFT, cyclic gates, phase gates, and quantum multiplexers, we develop the VBQC for Hamiltonian simulation (i.e. computing ${{e}^{\text{i}\mathbf{A}t}}$) and design the corresponding quantum circuit. The computational complexity of the designed quantum circuit mainly depends on the QFT and quantum multiplexers. Under the premise that quantum multiplexers can be implemented in the future, the proposed Hamiltonian simulation method based on KCQ decomposition offers a computational complexity advantage of $\text{poly}\left( \log N \right)$.
		
		Furthermore, this paper conducted simulations of the constructed quantum circuits on classical computers and verified the correctness of the proposed VBQC for Hamiltonian simulation based on the KCQ decomposition method through three numerical experiments. The study further combined the Hamiltonian simulation algorithm with quantum phase estimation to compute the free vibration frequencies and mode shapes of 1-dimensional, 2-dimensional, and 3-dimensional solid mechanics problems, comparing the results with classical finite element solutions to further validate the correctness of the proposed VBQC for solid mechanics and to demonstrate the accuracy of the proposed method for calculating the considered free vibrations problems. In the future, the research on quantum finite elements based on voxel grids will be further expanded, aiming to apply the constructed Hamiltonian simulation algorithm to the HHL algorithm and Schrödingerization method for investigating quantum simulation approaches for static and dynamic problems in solid mechanics. Meanwhile, we will further attempt to integrate VBQC with methods such as the implicit boundary method to tackle solid mechanics problems with complex boundary conditions.
		
		\appendix
		\renewcommand\thefigure{\Alph{section}\arabic{figure}}    
		\section{Several Lemmas}
		\setcounter{figure}{0} 
		\label{APPA}
		\noindent\textbf{Lemma 1} If a Hamiltonian $\mathbf{A}\in {{\mathbb{C}}^{{{2}^{n}}\times {{2}^{n}}}}$ can be decomposed into multiple subterms $\mathbf{A}=\sum\limits_{k=1}^{K}{{{\mathbf{A}}_{k}}}$, then
		\begin{equation}
			\begin{aligned}
				{{e}^{\text{i}\mathbf{A}\Delta t}} & =\prod\limits_{k=1}^{K}{{{e}^{\text{i}{{\mathbf{A}}_{k}}\Delta t}}}+O\left( \Delta t \right)={{\mathbf{S}}_{2}}\left( \Delta t \right)+O\left( \Delta {{t}^{2}} \right) \\ & ={{\mathbf{S}}_{2m}}\left( \Delta t \right)+O\left( \Delta {{t}^{2m+1}} \right)
			\end{aligned},
			\label{AE1}
		\end{equation}
		where
		\begin{equation}
			\begin{aligned}
				& {{\mathbf{S}}_{2}}\left( \Delta t \right)=\prod\limits_{k=1}^{K}{{{e}^{\text{i}{{\mathbf{A}}_{k}}{\Delta t}/{2}\;}}}\prod\limits_{k=K}^{1}{{{e}^{\text{i}{{\mathbf{A}}_{k}}{\Delta t}/{2}\;}}}, \\ 
				& {{\mathbf{S}}_{2m}}\left( \Delta t \right)={{\left[ {{\mathbf{S}}_{2m-2}}\left( {{p}_{m}}\Delta t \right) \right]}^{2}}{{\mathbf{S}}_{2m-2}}\left( \left( 1-4{{p}_{m}} \right)\Delta t \right){{\left[ {{\mathbf{S}}_{2m-2}}\left( {{p}_{m}}\Delta t \right) \right]}^{2}}, \\ 
				& {{p}_{m}}={{\left( 4-{{4}^{{1}/{\left( 2m-1 \right)}\;}} \right)}^{-1}},\ \ \ m>1.
			\end{aligned}
			\label{AE2}
		\end{equation}

		This lemma is adapted from the literature \citep{R20}. According to Lemma 1, if ${{e}^{\text{i}{{\mathbf{A}}_{k}}\Delta t}}$ can be easily simulated and $K\ll {{2}^{n}}$, then ${{e}^{\text{i}\mathbf{A}\Delta t}}$ can be effectively simulated using Eq. (\ref{AE1}).
		
		\noindent\textbf{Lemma 2} A cyclic shift matrix defined as ${{\mathbf{S}}_{N}}=\left[ \begin{matrix}
			\mathbf{0} & {{\mathbf{I}}_{N-1}}  \\
			1 & \mathbf{0}  \\
		\end{matrix} \right]$ can be decomposed as
		\begin{equation}
			{{\mathbf{S}}_{N}}={{\mathbf{Q}}_{N}}{{\mathbf{\Lambda }}_{N}}\mathbf{Q}_{N}^{\text{H}},
			\label{AE3}
		\end{equation}
		where
		\begin{equation}
			\begin{aligned}
				& {{\mathbf{\Lambda }}_{N}}=\underset{k=0}{\overset{N-1}{\mathop{\oplus }}}\,{{e}^{\frac{2\pi \text{i}}{N}k}}=\underset{k=n}{\overset{1}{\mathop{\otimes }}}\,\mathbf{T}\left( \frac{{{2}^{k}}\pi }{N} \right)=\underset{k=n}{\overset{1}{\mathop{\otimes }}}\,\left[ \begin{matrix}
					1 & 0  \\
					0 & {{e}^{\frac{{{2}^{k}}\pi \text{i}}{N}}}  \\
				\end{matrix} \right], \\ 
				& {{\mathbf{Q}}_{N}}=\frac{1}{\sqrt{N}}\left[ \begin{matrix}
					1 & 1 & 1 & \cdots  & 1  \\
					1 & {{e}^{\frac{2\pi \text{i}}{N}1}} & {{e}^{\frac{2\pi \text{i}}{N}2}} & \cdots  & {{e}^{\frac{2\pi \text{i}}{N}\left( N-1 \right)}}  \\
					1 & {{e}^{\frac{2\pi \text{i}}{N}2\times 1}} & {{e}^{\frac{2\pi \text{i}}{N}2\times 2}} & \cdots  & {{e}^{\frac{2\pi \text{i}}{N}2\left( N-1 \right)}}  \\
					\vdots  & \vdots  & \vdots  & \ddots  & \vdots   \\
					1 & {{e}^{\frac{2\pi \text{i}}{N}\left( N-1 \right)\times 1}} & {{e}^{\frac{2\pi \text{i}}{N}\left( N-1 \right)\times 2}} & \cdots  & {{e}^{\frac{2\pi \text{i}}{N}\left( N-1 \right)\left( N-1 \right)}}  \\
				\end{matrix} \right]. 
			\end{aligned}
			\label{AE4}
		\end{equation}
		
		\noindent\textbf{Proof:} It is necessary to prove that the $k\text{-th}$ column vector of ${{\mathbf{Q}}_{N}}$ is an eigenvector of ${{\mathbf{S}}_{N}}$. Obviously,	
		\begin{equation}
			\begin{aligned}
				& \left[ \begin{matrix}
					{} & 1 & {} & {} & {}  \\
					{} & {} & 1 & {} & {}  \\
					{} & {} & {} & \ddots  & {}  \\
					{} & {} & {} & {} & 1  \\
					1 & {} & {} & {} & {}  \\
				\end{matrix} \right]\left( \begin{matrix}
					1  \\
					{{e}^{\frac{2\pi \text{i}}{N}k}}  \\
					{{e}^{\frac{2\pi \text{i}}{N}2\times k}}  \\
					\vdots   \\
					{{e}^{\frac{2\pi \text{i}}{N}\left( N-1 \right)\times k}}  \\
				\end{matrix} \right)=\left( \begin{matrix}
					{{e}^{\frac{2\pi \text{i}}{N}k}}  \\
					{{e}^{\frac{2\pi \text{i}}{N}2\times k}}  \\
					\vdots   \\
					{{e}^{\frac{2\pi \text{i}}{N}\left( N-1 \right)\times k}}  \\
					1  \\
				\end{matrix} \right) \\ & ={{e}^{\frac{2\pi \text{i}}{N}k}}\left( \begin{matrix}
					1  \\
					{{e}^{\frac{2\pi \text{i}}{N}k}}  \\
					\vdots   \\
					{{e}^{\frac{2\pi \text{i}}{N}\left( N-2 \right)\times k}}  \\
					{{e}^{-\frac{2\pi \text{i}}{N}k}}  \\
				\end{matrix} \right)
			\end{aligned}.
			\label{AE5}
		\end{equation}
		Since
		\begin{equation}
			{{e}^{-\frac{2\pi \text{i}}{N}k}}={{e}^{\frac{2\pi \text{i}}{N}Nk}}{{e}^{-\frac{2\pi \text{i}}{N}k}}={{e}^{\frac{2\pi \text{i}}{N}\left( N\times k-k \right)}}={{e}^{\frac{2\pi \text{i}}{N}\left( N-1 \right)k}},
			\label{AE6}
		\end{equation}
		the $k\text{-th}$ column vector of ${{\mathbf{Q}}_{N}}$ is an eigenvector of ${{\mathbf{S}}_{N}}$, with the eigenvalue ${{e}^{\frac{2\pi \text{i}}{N}k}}$. $\blacksquare$
						
\begin{figure}[!h]
			\centering
			\includegraphics{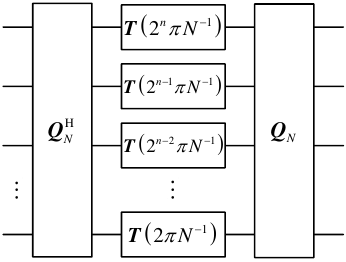}
			\caption{Quantum circuit for the cyclic matrix.}
			\label{AF1}
		\end{figure}

		It can be seen that ${{\mathbf{Q}}_{N}}$ is the Fourier transform matrix, which can be solved using the QFT method. Therefore, the quantum circuit for the cyclic matrix ${{\mathbf{S}}_{N}}$ is shown in Fig. \ref{AF1}.
		
		\noindent\textbf{Lemma 3} If the matrix $\mathbf{A}=\left[ \begin{matrix}
			a & b  \\
			b & c  \\
		\end{matrix} \right]$ is a real Hermitian matrix, we have
		\begin{equation}
			{{e}^{\text{i}\mathbf{A}}}=\mathbf{R}\left( \theta \left( \mathbf{A} \right) \right)\mathbf{T}\left( {{\lambda }_{2}}\left( \mathbf{A} \right) \right)\mathbf{XT}\left( {{\lambda }_{1}}\left( \mathbf{A} \right) \right)\mathbf{X}{{\mathbf{R}}^{\text{H}}}\left( \theta \left( \mathbf{A} \right) \right),
			\label{AE7}
		\end{equation}
		where
		\begin{equation}
			\begin{aligned}
				& {{\lambda }_{1}}\left( \mathbf{A} \right)=\frac{a+c}{2}-\frac{\sqrt{{{\left( a-c \right)}^{2}}+4{{b}^{2}}}}{2},\\
				& {{\lambda }_{2}}\left( \mathbf{A} \right)=\frac{a+c}{2}+\frac{\sqrt{{{\left( a-c \right)}^{2}}+4{{b}^{2}}}}{2}, \\ 
				& \cos \left[ \theta \left( \mathbf{A} \right) \right]=\frac{\left| b \right|}{\sqrt{{{\left( {{\lambda }_{1}}-c \right)}^{2}}+{{b}^{2}}}}\frac{{{\lambda }_{1}}-c}{b}, \\
				& \sin \left[ \theta \left( \mathbf{A} \right) \right]=\frac{\left| b \right|}{\sqrt{{{\left( {{\lambda }_{1}}-c \right)}^{2}}+{{b}^{2}}}}, \\ 
				& \theta \left( \mathbf{A} \right)=-\text{i}\log \left[ \frac{\left| b \right|\left( {{\lambda }_{1}}-c+b\text{i} \right)}{b\sqrt{{{\left( {{\lambda }_{1}}-c \right)}^{2}}+{{b}^{2}}}} \right].
			\end{aligned}
			\label{AE8}
		\end{equation}
		
		\noindent\textbf{Proof:} Since$\mathbf{A}$ can be diagonalized as,
		\begin{equation}
			\mathbf{A}=\mathbf{R}\left( \theta  \right)\mathbf{D}{{\mathbf{R}}^{\text{H}}}\left( \theta  \right),
			\label{AE9}
		\end{equation}
		where
		\begin{equation}
			\begin{aligned}
				& \mathbf{D}=\left( \begin{matrix}
					{{\lambda }_{1}}\left( \mathbf{A} \right) & 0  \\
					0 & {{\lambda }_{2}}\left( \mathbf{A} \right)  \\
				\end{matrix} \right), \\ 
				& {{\lambda }_{1}}=\frac{a+c}{2}-\frac{\sqrt{{{\left( a-c \right)}^{2}}+4{{b}^{2}}}}{2},\ \ \ {{\lambda }_{2}}=\frac{a+c}{2}+\frac{\sqrt{{{\left( a-c \right)}^{2}}+4{{b}^{2}}}}{2}, \\ 
				& \cos \left( \theta \left( \mathbf{A} \right) \right)=\frac{\left| b \right|}{\sqrt{{{\left( {{\lambda }_{2}}-c \right)}^{2}}+{{b}^{2}}}},\ \ \ \ \sin \left( \theta \left( \mathbf{A} \right) \right)=\frac{-\left| b \right|}{\sqrt{{{\left( {{\lambda }_{1}}-c \right)}^{2}}+{{b}^{2}}}}. 
			\end{aligned}
			\label{AE10}
		\end{equation}
		Therefore
		\begin{equation}
			\begin{aligned}
				{{e}^{\text{i}\mathbf{A}}} & ={{e}^{\text{i}\mathbf{R}\left( \theta  \right)\mathbf{D}{{\mathbf{R}}^{\text{H}}}\left( \theta  \right)}}=\mathbf{R}\left( \theta  \right)\left( \begin{matrix}
					{{e}^{\text{i}{{\lambda }_{1}}\left( \mathbf{A} \right)}} & 0  \\
					0 & {{e}^{\text{i}{{\lambda }_{2}}\left( \mathbf{A} \right)}}  \\
				\end{matrix} \right){{\mathbf{R}}^{\text{H}}}\left( \theta  \right) \\ 
				& =\mathbf{R}\left( \theta \left( \mathbf{A} \right) \right)\mathbf{T}\left( {{\lambda }_{2}}\left( \mathbf{A} \right) \right)\mathbf{XT}\left( {{\lambda }_{1}}\left( \mathbf{A} \right) \right)\mathbf{X}{{\mathbf{R}}^{\text{H}}}\left( \theta \left( \mathbf{A} \right) \right) 
			\end{aligned},
			\label{AE11}
		\end{equation}
		which completes the proof. $\blacksquare$
		
		\noindent\textbf{Lemma 4} If the matrix $\mathbf{A}=\left[ \begin{matrix}
			0 & -b  \\
			b & 0  \\
		\end{matrix} \right]$ is a real skew-symmetric matrix, we have
		\begin{equation}
			{{e}^{\mathbf{A}}}={{e}^{\left[ \begin{matrix}
						0 & -b  \\
						b & 0  \\
					\end{matrix} \right]}}=\mathbf{R}\left( b \right)=\left[ \begin{matrix}
				\cos b & -\sin b  \\
				\sin b & \cos b  \\
			\end{matrix} \right].
			\label{AE12}
		\end{equation}
				
		\noindent\textbf{Proof:} Since $\mathbf{A}$ can be diagonalized as,
		\begin{equation}
			\mathbf{A}=\mathbf{\Phi D}{{\mathbf{\Phi }}^{\text{H}}},\ \ \mathbf{D}=\left[ \begin{matrix}
				-b\text{i} & 0  \\
				0 & b\text{i}  \\
			\end{matrix} \right],\ \ \mathbf{\Phi }=\frac{1}{\sqrt{2}}\left[ \begin{matrix}
				-\text{i} & \text{i}  \\
				1 & 1  \\
			\end{matrix} \right].
			\label{AE13}
		\end{equation}
		
		Therefore,
		\begin{equation}
			{{e}^{\mathbf{A}}}={{e}^{\mathbf{\Phi D}{{\mathbf{\Phi }}^{\text{H}}}}}=\frac{1}{2}\left[ \begin{matrix}
				-\text{i} & \text{i}  \\
				1 & 1  \\
			\end{matrix} \right]\left[ \begin{matrix}
				{{e}^{-b\text{i}}} & 0  \\
				0 & {{e}^{b\text{i}}}  \\
			\end{matrix} \right]\left[ \begin{matrix}
				\text{i} & 1  \\
				-\text{i} & 1  \\
			\end{matrix} \right]=\left[ \begin{matrix}
				\cos b & -\sin b  \\
				\sin b & \cos b  \\
			\end{matrix} \right],
			\label{AE14}
		\end{equation}
		which completes the proof. $\blacksquare$
		
		\section{Quantum Circuits for Matrix Exponentials of Low-Order Hermitian Matrices}
		\label{APPB}
		\setcounter{figure}{0} 
		Assume there is a $d\text{-dimensional}$ Hermitian matrix $\mathbf{A}$, denoted by
		\begin{equation}
			\mathbf{A}=\left[ \begin{matrix}
				{{\mathbf{a}}_{11}} & {{\mathbf{a}}_{12}}  \\
				{{\mathbf{a}}_{21}} & {{\mathbf{a}}_{22}}  \\
			\end{matrix} \right].
			\label{BE1}
		\end{equation}
		The quantum circuit for the matrix exponential ${{e}^{\text{i}\Delta t\mathbf{A}}}$ is denoted as ${{\mathbf{W}}_{d}}\left( \mathbf{A}\Delta t \right)$. When $d=2$ and $\mathbf{A}$ is a real symmetric matrix, according to Lemma 3, it follows that:
		\begin{equation}
			{{\mathbf{W}}_{2}}\left( \mathbf{A}\Delta t \right):={{e}^{\text{i}\Delta t\mathbf{A}}}=\mathbf{R}\left( \theta \left( \mathbf{A} \right) \right)\mathbf{T}\left( \Delta t{{\lambda }_{2}}\left( \mathbf{A} \right) \right)\mathbf{XT}\left( \Delta t{{\lambda }_{1}}\left( \mathbf{A} \right) \right)\mathbf{X}{{\mathbf{R}}^{\text{H}}}\left( \theta \left( \mathbf{A} \right) \right).
			\label{BE2}
		\end{equation}
		
		When $d=2$ and $\mathbf{A}$ is purely imaginary, its form must be:
		\begin{equation}
			\mathbf{A}=\left[ \begin{matrix}
				0 & \text{i}a  \\
				-\text{i}a & 0  \\
			\end{matrix} \right]
			\label{BE3}
		\end{equation}
		And according to Lemma 4, it can be obtained
		\begin{equation}
			{{\mathbf{W}}_{2}}\left( \mathbf{A}\Delta t \right):={{e}^{\text{i}\Delta t\mathbf{A}}}={{e}^{\text{i}\Delta t\left[ \begin{matrix}
						0 & \text{i}a  \\
						-\text{i}a & 0  \\
					\end{matrix} \right]}}=\mathbf{R}\left( a\Delta t \right).
			\label{BE4}
		\end{equation}
		
\begin{figure}[!h]
			\centering
			\includegraphics[width=1\textwidth]{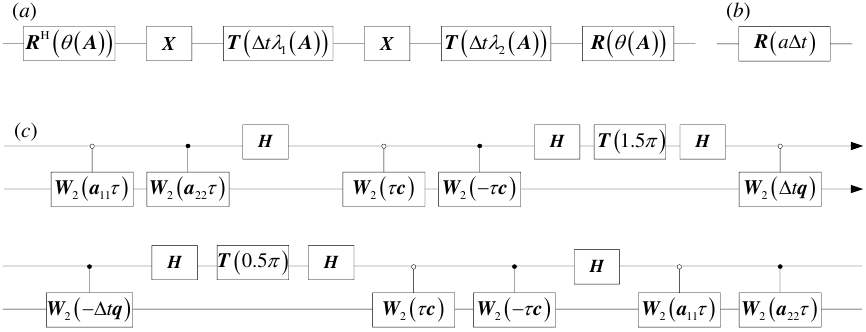}
			\caption{Quantum circuits for low-order Hermitian matrices: (a) The quantum circuit for Eq. (\ref{BE2}); (b) The quantum circuit for Eq. (\ref{BE4}); (c) The quantum circuit for ${{\mathbf{W}}_{4}}\left( \mathbf{A}\Delta t \right)$.}
			\label{BF1}
		\end{figure}

		When $d=4$, using the KCQ decomposition, Eq. (\ref{BE1}) can be decomposed into
		\begin{equation}
			\begin{aligned}
				& \mathbf{A}={{\mathbf{A}}_{1}}+{{\mathbf{A}}_{2}}+{{\mathbf{A}}_{3}}, \\ 
				& {{\mathbf{A}}_{1}}={{\mathbf{a}}_{11}}\oplus {{\mathbf{a}}_{22}}, \\ 
				& {{\mathbf{A}}_{2}}=\left( \mathbf{H}\otimes {{\mathbf{I}}_{2}} \right)\left[ \mathbf{c}\oplus \left( -\mathbf{c} \right) \right]\left( \mathbf{H}\otimes {{\mathbf{I}}_{2}} \right)\, \\ 
				& {{\mathbf{A}}_{3}}=\left( \mathbf{T}\left( 0.5\pi  \right)\mathbf{H}\otimes {{\mathbf{I}}_{2}} \right)\left[ \mathbf{q}\oplus \left( -\mathbf{q} \right) \right]\left( \mathbf{HT}\left( 1.5\pi  \right)\otimes {{\mathbf{I}}_{2}} \right), \\ 
			\end{aligned}
			\label{BE5}
		\end{equation}
		where
		\begin{equation}
			\mathbf{c}\left( \mathbf{A} \right)=\frac{1}{2}\left( {{\mathbf{a}}_{12}}+{{\mathbf{a}}_{21}} \right),\ \ \ \mathbf{q}\left( \mathbf{A} \right)=\frac{\text{i}}{2}\left( {{\mathbf{a}}_{12}}-{{\mathbf{a}}_{21}} \right).
			\label{BE6}
		\end{equation}
		It can be seen that ${{\mathbf{a}}_{ii}}$, $\mathbf{c}$, and $\mathbf{q}$ are all $2\times 2$ Hermitian matrices. The matrix exponential of each matrix ${{\mathbf{A}}_{k}}$ are
		\begin{equation}
			{{e}^{\text{i}\Delta t{{\mathbf{A}}_{1}}}}=\exp \left( \Delta t{{\mathbf{a}}_{11}} \right)\oplus \exp \left( \Delta t{{\mathbf{a}}_{22}} \right)={{\mathbf{W}}_{2}}\left( {{\mathbf{a}}_{11}}\Delta t \right)\oplus {{\mathbf{W}}_{2}}\left( {{\mathbf{a}}_{22}}\Delta t \right),
			\label{BE7}
		\end{equation}
		\begin{equation}
			{{e}^{\text{i}\Delta t{{\mathbf{A}}_{2}}}}=\left( \mathbf{H}\otimes {{\mathbf{I}}_{2}} \right)\left[ {{\mathbf{W}}_{2}}\left( \Delta t\mathbf{c} \right)\oplus {{\mathbf{W}}_{2}}\left( -\Delta t\mathbf{c} \right) \right]\left( \mathbf{H}\otimes {{\mathbf{I}}_{2}} \right),
			\label{BE8}
		\end{equation}
		\begin{equation}
			\begin{aligned}
				{{e}^{\text{i}\Delta t{{\mathbf{A}}_{3}}}} & =\left( \mathbf{T}\left( 0.5\pi  \right)\mathbf{H}\otimes {{\mathbf{I}}_{2}} \right)\left[ {{\mathbf{W}}_{2}}\left( \Delta t\mathbf{q} \right)\oplus {{\mathbf{W}}_{2}}\left( -\Delta t\mathbf{q} \right) \right] \\
				& \times {{\left( \mathbf{T}\left( 0.5\pi  \right)\mathbf{H}\otimes {{\mathbf{I}}_{2}} \right)}^{\text{H}}}
			\end{aligned},
			\label{BE9}
		\end{equation}
		According to Lemma 1, ${{\mathbf{W}}_{4}}\left( \mathbf{A}\Delta t \right)$ can be written as
		\begin{equation}
			\begin{aligned}
				{{\mathbf{W}}_{4}}\left( \mathbf{A}\Delta t \right) & =\left[ {{\mathbf{W}}_{2}}\left( {{\mathbf{a}}_{11}}\tau  \right)\oplus {{\mathbf{W}}_{2}}\left( {{\mathbf{a}}_{22}}\tau  \right) \right]\left( \mathbf{H}\otimes {{\mathbf{I}}_{2}} \right)\\ 
				&\times \left[ {{\mathbf{W}}_{2}}\left( \tau \mathbf{c} \right)\oplus {{\mathbf{W}}_{2}}\left( -\tau \mathbf{c} \right) \right]\left( \mathbf{H}\otimes {{\mathbf{I}}_{2}} \right)\left[ \left( \mathbf{T}\left( 0.5\pi  \right)\mathbf{H} \right)\otimes {{\mathbf{I}}_{2}} \right]\\ 
				&\times \left[ {{\mathbf{W}}_{2}}\left( \Delta t\mathbf{q} \right)\oplus {{\mathbf{W}}_{2}}\left( -\Delta t\mathbf{q} \right) \right]\left[ \left( \mathbf{HT}\left( 1.5\pi  \right) \right)\otimes {{\mathbf{I}}_{2}} \right]\\
				&\times \left( \mathbf{H}\otimes {{\mathbf{I}}_{2}} \right)\left[ {{\mathbf{W}}_{2}}\left( \tau \mathbf{c} \right)\oplus {{\mathbf{W}}_{2}}\left( -\tau \mathbf{c} \right) \right]\left( \mathbf{H}\otimes {{\mathbf{I}}_{2}} \right)\\
				&\times \left[ {{\mathbf{W}}_{2}}\left( {{\mathbf{a}}_{11}}\tau  \right)\oplus {{\mathbf{W}}_{2}}\left( {{\mathbf{a}}_{22}}\tau  \right) \right]+O\left( \Delta {{t}^{3}} \right) \\ 
			\end{aligned},
			\label{BE10}
		\end{equation}
		where $\tau ={\Delta t}/{2}\;$. The quantum circuits of Eqs. (\ref{BE2}), (\ref{BE4}), and(\ref{BE10}) are shown in Fig. \ref{BF1}.
		
		\section{Quantum Phase Estimation Algorithm}
		\label{APPC} 
		\setcounter{figure}{0} 
		\begin{figure}[!h]
			\centering
			\includegraphics[width=1\textwidth]{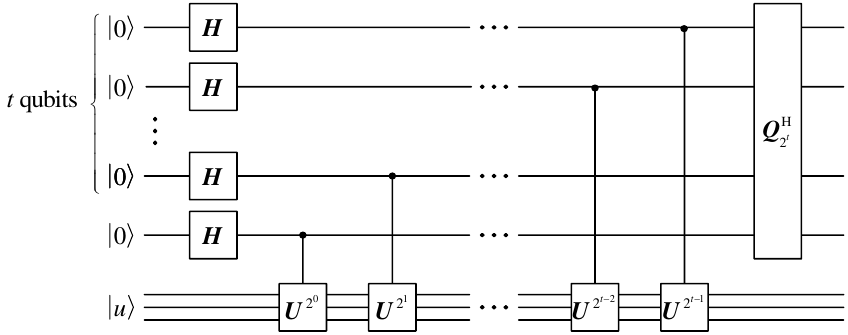}
			\caption{Quantum circuit for Quantum Phase Estimation Algorithm.}
			\label{CF1}
		\end{figure}
		For a Hermitian matrix $\mathbf{\bar{K}}$ and a quantum state $\left| u \right\rangle =\sum\limits_{i=0}^{N-1}{{{u}_{i}}\left| {{x}_{i}} \right\rangle }$ ,where $\left| {{x}_{i}} \right\rangle$ are eigenstates of $\mathbf{\bar{K}}$. Let $\mathbf{U}={{e}^{\text{i}\mathbf{\bar{K}}\Delta t}}$, applying quantum phase estimation to ${{\left| 0 \right\rangle }^{\otimes t}}\left| u \right\rangle $ yields
		\begin{equation}
			{{\left| 0 \right\rangle }^{\otimes t}}\left| u \right\rangle \xrightarrow{\text{QPE}}\sum\limits_{i=0}^{N-1}{{{u}_{i}}\left| {{x}_{i}} \right\rangle \left| {{{\bar{\varphi }}}_{i}} \right\rangle },
			\label{CE1}
		\end{equation}
		where $\left| {{{\bar{\varphi }}}_{i}} \right\rangle $ is an approximation of ${\Delta t{{\lambda }_{i}}}/{2\pi }\;$, with ${{\lambda }_{i}}$ being the eigenvalue of $\mathbf{\bar{K}}$ and $t$ being the number of auxiliary qubits. The quantum circuit diagram of the quantum phase estimation algorithm is shown in Fig. \ref{CF1}. For detailed information on the quantum phase estimation algorithm, readers can refer to the literature \citep{R28}.

		\section*{Acknowledgments}
		The authors are grateful for the support of the Natural Science Foundation of China (No. 12372190).

		\bibliographystyle{elsarticle-num-names} 
		\bibliography{Reference}

@article{R1,
  title={Hybrid quantum physics-informed neural network: Towards efficient learning of high-speed flows},
  author={Leong, Fong Yew and Ewe, Wei-Bin and Quang, Tran Si Bui and Zhang, Zhong Yuan and Khoo, Jun Yong},
  journal={arXiv preprint arXiv:2503.02202},
  year={2025}
}

@article{R2,
  title={Quantum simulation of partial differential equations: Applications and detailed analysis},
  author={Jin, Shi and Liu, Na Na and Yu, Yue},
  journal={Physical Review A},
  volume={108},
  number={3},
  pages={032603},
  year={2023},
  publisher={APS}
}

@article{R3,
  title={Quantum Simulation of Partial Differential Equations via Schr{\"o}dingerization},
  author={Jin, Shi and Liu, Na Na and Yu, Yue},
  journal={Physical Review Letters},
  volume={133},
  number={23},
  pages={230602},
  year={2024},
  publisher={APS}
}

@article{R4,
  title={Quantum computing of fluid dynamics using the hydrodynamic Schr{\"o}dinger equation},
  author={Meng, Zhao Yuan and Yang, Yue},
  journal={Physical Review Research},
  volume={5},
  number={3},
  pages={033182},
  year={2023},
  publisher={APS}
}

@article{R5,
  title={Quantum computing of reacting flows via Hamiltonian simulation},
  author={Lu, Zhen and Yang, Yue},
  journal={Proceedings of the Combustion Institute},
  volume={40},
  number={1-4},
  pages={105440},
  year={2024},
  publisher={Elsevier}
}

@article{R6,
  title={Simulating unsteady flows on a superconducting quantum processor},
  author={Meng, Zhao Yuan and Zhong, Jia Run and Xu, Shi Bo and Wang, Ke and Chen, Jia Chen and Jin, Fei Tong and Zhu, Xu Hao and Gao, Yu and Wu, Yao Zu and Zhang, Chuan Yu and others},
  journal={Communications Physics},
  volume={7},
  number={1},
  pages={349},
  year={2024},
  publisher={Nature Publishing Group UK London}
}

@article{R7,
  title={Applications of the vortex-surface field to flow visualization, modelling and simulation},
  author={Yang, Yue and Xiong, Shi Ying and Lu, Zhen},
  journal={Flow},
  volume={3},
  pages={E33},
  year={2023},
  publisher={Cambridge University Press}
}

@article{R8,
  title={Quantum algorithm for linear systems of equations},
  author={Harrow, Aram W and Hassidim, Avinatan and Lloyd, Seth},
  journal={Physical review letters},
  volume={103},
  number={15},
  pages={150502},
  year={2009},
  publisher={APS}
}

@article{R9,
  title={Finding flows of a Navier--Stokes fluid through quantum computing},
  author={Gaitan, Frank},
  journal={npj Quantum Information},
  volume={6},
  number={1},
  pages={61},
  year={2020},
  publisher={Nature Publishing Group UK London}
}

@article{R10,
  title={Hybrid quantum algorithms for flow problems},
  author={Bharadwaj, Sachin S and Sreenivasan, Katepalli R},
  journal={Proceedings of the National Academy of Sciences},
  volume={120},
  number={49},
  pages={e2311014120},
  year={2023},
  publisher={National Academy of Sciences}
}

@article{R11,
  title={Quantum computing for solid mechanics and structural engineering--A demonstration with Variational Quantum Eigensolver},
  author={Liu, Yun Ya and Liu, Jia Kun and Raney, Jordan R and Wang, Pai},
  journal={Extreme Mechanics Letters},
  volume={67},
  pages={102117},
  year={2024},
  publisher={Elsevier}
}

@article{R12,
  title={Improved circuit implementation of the HHL algorithm and its simulations on QISKIT},
  author={Zhang, Meng and Dong, Li Hua and Zeng, Yong and Cao, Ning},
  journal={Scientific Reports},
  volume={12},
  number={1},
  pages={13287},
  year={2022},
  publisher={Nature Publishing Group UK London}
}

@article{R13,
  title={An implementation of the finite element method in hybrid classical/quantum computers},
  author={Arora, Abhishek and Ward, Benjamin M and Oskay, Caglar},
  journal={Finite Elements in Analysis and Design},
  volume={248},
  pages={104354},
  year={2025},
  publisher={Elsevier}
}

@article{R14,
  title={A quantum annealing-sequential quadratic programming assisted finite element simulation for non-linear and history-dependent mechanical problems},
  author={Nguyen, Van-Dung and Wu, Ling and Remacle, Fran{\c{c}}oise and Noels, Ludovic},
  journal={European Journal of Mechanics-A/Solids},
  volume={105},
  pages={105254},
  year={2024},
  publisher={Elsevier}
}

@article{R15,
  title={A novel design update framework for topology optimization with quantum annealing: Application to truss and continuum structures},
  author={Sukulthanasorn, Naruethep and Xiao, Jun Sen and Wagatsuma, Koya and Nomura, Reika and Moriguchi, Shuji and Terada, Kenjiro},
  journal={Computer Methods in Applied Mechanics and Engineering},
  volume={437},
  pages={117746},
  year={2025},
  publisher={Elsevier}
}

@article{R16,
  title={Parallel evaluation of quantum algorithms for computational fluid dynamics},
  author={Steijl, Ren{\'e} and Barakos, George N},
  journal={Computers \& Fluids},
  volume={173},
  pages={22--28},
  year={2018},
  publisher={Elsevier}
}

@article{R17,
  title={Solving linear systems on quantum hardware with hybrid HHL++},
  author={Yalovetzky, Romina and Minssen, Pierre and Herman, Dylan and Pistoia, Marco},
  journal={Scientific Reports},
  volume={14},
  number={1},
  pages={20610},
  year={2024},
  publisher={Nature Publishing Group UK London}
}

@article{R18,
  title={Variational quantum algorithms},
  author={Cerezo, Marco and Arrasmith, Andrew and Babbush, Ryan and Benjamin, Simon C and Endo, Suguru and Fujii, Keisuke and McClean, Jarrod R and Mitarai, Kosuke and Yuan, Xiao and Cincio, Lukasz and others},
  journal={Nature Reviews Physics},
  volume={3},
  number={9},
  pages={625--644},
  year={2021},
  publisher={Nature Publishing Group UK London}
}

@book{R19,
  title={The finite element method for solid and structural mechanics},
  author={Zienkiewicz, Olgierd Cecil and Taylor, Robert Leroy},
  year={2005},
  publisher={Elsevier}
}

@article{R20,
  title={Efficient quantum algorithms for simulating sparse Hamiltonians},
  author={Berry, Dominic W and Ahokas, Graeme and Cleve, Richard and Sanders, Barry C},
  journal={Communications in Mathematical Physics},
  volume={270},
  pages={359--371},
  year={2007},
  publisher={Springer}
}

@article{R21,
  title={Black-box hamiltonian simulation and unitary implementation},
  author={Berry, Dominic W and Childs, Andrew M},
  journal={Quantum Information and Computation},
  volume={12},
  number={1-2},
  pages={29--62},
  year={2012},
  publisher={Rinton Press Inc.}
}

@mastersthesis{R22,
  author  = {Ahokas, Graeme Robert},
  title   = {Improved algorithms for approximate quantum Fourier transforms and sparse Hamiltonian simulations},
  school  = {University of Calgar},
  year    = {2004}
}

@article{R23,
  title={Finite difference method for numerical computation of discontinuous solutions of the equations of fluid dynamics},
  author={Godunov, Sergei K and Bohachevsky, Ihor},
  journal={Matemati{\v{c}}eskij sbornik},
  volume={47},
  number={3},
  pages={271--306},
  year={1959}
}

@article{R24,
  title={The finite difference method in partial differential equations},
  author={Mitchell, Andrew Ronald and Griffiths, David Francis},
  journal={A Wiley-Interscience Publication},
  year={1980}
}

@article{R25,
  title={A mechanics-based data-free Problem Independent Machine Learning (PIML) model for large-scale structural analysis and design optimization},
  author={Huang, Meng Cheng and Liu, Chang and Guo, Yi Lin and Zhang, Lin Feng and Du, Zong Liang and Guo, Xu},
  journal={Journal of the Mechanics and Physics of Solids},
  volume={193},
  pages={105893},
  year={2024},
  publisher={Elsevier}
}

@article{R26,
  title={A Problem-Independent Machine Learning (PIML) enhanced substructure-based approach for large-scale structural analysis and topology optimization of linear elastic structures},
  author={Huang, Meng Cheng and Cui, Tian Chen and Liu, Chang and Du, Zong Liang and Zhang, Jia Meng and He, Chu Hui and Guo, Xu},
  journal={Extreme Mechanics Letters},
  volume={63},
  pages={102041},
  year={2023},
  publisher={Elsevier}
}

@article{R27,
  title={Problem-independent machine learning (PIML)-based topology optimization—A universal approach},
  author={Huang, Meng Cheng and Du, Zong Liang and Liu, Chang and Zheng, Yong Gang and Cui, Tian Chen and Mei, Yue and Li, Xiao and Zhang, Xiao Yu and Guo, Xu},
  journal={Extreme Mechanics Letters},
  volume={56},
  pages={101887},
  year={2022},
  publisher={Elsevier}
}

@book{R28,
  title={Quantum computation and quantum information},
  author={Nielsen, Michael A and Chuang, Isaac L},
  year={2010},
  publisher={Cambridge university press}
}

@phdthesis{R29,
  author  = {Chen, Jia Lin},
  title   = {Programmable architecture research of quantum computing},
  school  = {Fu Dan University},
  year    = {2013}
}

@article{R30,
  title={Quantum Fredkin and Toffoli gates on a versatile programmable silicon photonic chip},
  author={Li, Yuan and Wan, Ling Xiao and Zhang, Hui and Zhu, Hui Hui and Shi, Yu Zhi and Chin, Lip Ket and Zhou, Xiao Qi and Kwek, Leong Chuan and Liu, Ai Qun},
  journal={npj Quantum Information},
  volume={8},
  number={1},
  pages={112},
  year={2022},
  publisher={Nature Publishing Group UK London}
}

@article{R31,
  title={Polylogarithmic-depth controlled-NOT gates without ancilla qubits},
  author={Claudon, Baptiste and Zylberman, Julien and Feniou, C{\'e}sar and Debbasch, Fabrice and Peruzzo, Alberto and Piquemal, Jean-Philip},
  journal={Nature Communications},
  volume={15},
  number={1},
  pages={5886},
  year={2024},
  publisher={Nature Publishing Group UK London}
}

@article{R32,
  title={Characterization, synthesis, and optimization of quantum circuits over multiple-control Z-rotation gates: A systematic study},
  author={Zhang, Shi Hao and Wu, Jun Da and Li, Lv Zhou},
  journal={Physical Review A},
  volume={108},
  number={2},
  pages={022603},
  year={2023},
  publisher={APS}
}

@article{R33,
  title={Hardware-efficient quantum random access memory with hybrid quantum acoustic systems},
  author={Hann, Connor T and Zou, Chang-Ling and Zhang, Ya Xing and Chu, Yi Wen and Schoelkopf, Robert J and Girvin, Steven M and Jiang, Liang},
  journal={Physical review letters},
  volume={123},
  number={25},
  pages={250501},
  year={2019},
  publisher={APS}
}

@article{R34,
  title={Circuit-based quantum random access memory for classical data},
  author={Park, Daniel K and Petruccione, Francesco and Rhee, June-Koo Kevin},
  journal={Scientific reports},
  volume={9},
  number={1},
  pages={3949},
  year={2019},
  publisher={Nature Publishing Group UK London}
}

@article{R35,
  title={Quantum random access memory},
  author={Giovannetti, Vittorio and Lloyd, Seth and Maccone, Lorenzo},
  journal={Physical review letters},
  volume={100},
  number={16},
  pages={160501},
  year={2008},
  publisher={APS}
}

@article{R36,
  title={Quantum random access memory architectures using 3D superconducting cavities},
  author={Weiss, DK and Puri, Shruti and Girvin, SM},
  journal={PRX Quantum},
  volume={5},
  number={2},
  pages={020312},
  year={2024},
  publisher={APS}
}

@article{R37,
  title={Observation of Rydberg blockade between two atoms},
  author={Urban, E and Johnson, Todd A and Henage, T and Isenhower, L and Yavuz, DD and Walker, TG and Saffman, M},
  journal={Nature Physics},
  volume={5},
  number={2},
  pages={110--114},
  year={2009},
  publisher={Nature Publishing Group UK London}
}

@article{R38,
  title={Demonstration of fault-tolerant universal quantum gate operations},
  author={Postler, Lukas and Heu$\beta$en, Sascha and Pogorelov, Ivan and Rispler, Manuel and Feldker, Thomas and Meth, Michael and Marciniak, Christian D and Stricker, Roman and Ringbauer, Martin and Blatt, Rainer and others},
  journal={Nature},
  volume={605},
  number={7911},
  pages={675--680},
  year={2022},
  publisher={Nature Publishing Group UK London}
}

@article{R39,
  title={Demonstration of a quantum gate using electromagnetically induced transparency},
  author={McDonnell, Katie and Keary, LF and Pritchard, JD},
  journal={Physical Review Letters},
  volume={129},
  number={20},
  pages={200501},
  year={2022},
  publisher={APS}
}

@article{R40,
  title={Quantum algorithm for data fitting},
  author={Wiebe, Nathan and Braun, Daniel and Lloyd, Seth},
  journal={Physical review letters},
  volume={109},
  number={5},
  pages={050505},
  year={2012},
  publisher={APS}
}

@article{R41,
  title={Quantum support vector machine for big data classification},
  author={Rebentrost, Patrick and Mohseni, Masoud and Lloyd, Seth},
  journal={Physical review letters},
  volume={113},
  number={13},
  pages={130503},
  year={2014},
  publisher={APS}
}

@article{R42,
  title={Quantum discriminant analysis for dimensionality reduction and classification},
  author={Cong, Iris and Duan, Lu Ming},
  journal={New Journal of Physics},
  volume={18},
  number={7},
  pages={073011},
  year={2016},
  publisher={IOP Publishing}
}

@article{R43,
  title={Quantum k-medoids algorithm using parallel amplitude estimation},
  author={Li, Yong-Mei and Liu, Hai-Ling and Pan, Shi-Jie and Qin, Su-Juan and Gao, Fei and Sun, Dong-Xu and Wen, Qiao-Yan},
  journal={Physical Review A},
  volume={107},
  number={2},
  pages={022421},
  year={2023},
  publisher={APS}
}

@article{R44,
  title={An adaptively filtered precise integration method considering perturbation for stochastic dynamics problems},
  author={Zhu, Li and Ye, Ke Qi and Huang, Dong Wei and Wu, Feng and Zhong, Wan Xie},
  journal={Acta Mechanica Solida Sinica},
  volume={36},
  number={2},
  pages={317--326},
  year={2023},
  publisher={Springer}
}

@article{R45,
  title={A precise time step integration method},
  author={Zhong, WX and Williams, FW},
  journal={Proceedings of the Institution of Mechanical Engineers, Part C: Journal of Mechanical Engineering Science},
  volume={208},
  number={6},
  pages={427--430},
  year={1994},
  publisher={SAGE Publications Sage UK: London, England}
}

@article{RBHu1,
title = {Quantum computing enhanced distance-minimizing data-driven computational mechanics},
journal = {Computer Methods in Applied Mechanics and Engineering},
volume = {419},
pages = {116675},
year = {2024},
issn = {0045-7825},
author = {Yongchun Xu and Jie Yang and Zengtao Kuang and Qun Huang and Wei Huang and Heng Hu},
}

@article{RBHu2,
title = {Quantum computing with error mitigation for data-driven computational homogenization},
journal = {Composite Structures},
volume = {351},
pages = {118625},
year = {2025},
issn = {0263-8223},
author = {Zengtao Kuang and Yongchun Xu and Qun Huang and Jie Yang and Chafik El Kihal and Heng Hu},
}

@article{RBVBG,
  title={The finite cell method for bone simulations: verification and validation},
  author={Ruess, Martin and Tal, David and Trabelsi, Nir and Yosibash, Zohar and Rank, Ernst},
  journal={Biomechanics and modeling in mechanobiology},
  volume={11},
  pages={425--437},
  year={2012},
  publisher={Springer}
}

@article{RBVBY,
  title={Numerical homogenization of heterogeneous and cellular materials utilizing the finite cell method},
  author={Düster, Alexander and Sehlhorst, Hans-Georg and Rank, Ernst},
  journal={Computational Mechanics},
  volume={50},
  pages={413--431},
  year={2012},
  publisher={Springer}
}

@article{RBFCM1,
  title={The finite cell method for three-dimensional problems of solid mechanics},
  author={D{\"u}ster, Alexander and Parvizian, Jamshid and Yang, Zhengxiong and Rank, Ernst},
  journal={Computer methods in applied mechanics and engineering},
  volume={197},
  number={45-48},
  pages={3768--3782},
  year={2008},
  publisher={Elsevier}
}

@article{RBFCM2,
  title={Finite cell method: h-and p-extension for embedded domain problems in solid mechanics},
  author={Parvizian, Jamshid and D{\"u}ster, Alexander and Rank, Ernst},
  journal={Computational Mechanics},
  volume={41},
  number={1},
  pages={121--133},
  year={2007},
  publisher={Springer}
}

@article{RBIBM1,
author = {Kumar, Ashok V. and Padmanabhan, Sanjeev and Burla, Ravi},
title = {Implicit boundary method for finite element analysis using non-conforming mesh or grid},
journal = {International Journal for Numerical Methods in Engineering},
volume = {74},
number = {9},
pages = {1421-1447},
year = {2008}
}

@article{RBIBM2,
  title={Implicit boundary method for analysis using uniform B-spline basis and structured grid},
  author={Burla, Ravi K and Kumar, Ashok V},
  journal={International journal for numerical methods in engineering},
  volume={76},
  number={13},
  pages={1993--2028},
  year={2008},
  publisher={Wiley Online Library}
}

@article{RBFBM,
  title={A fat boundary method for the Poisson problem in a domain with holes},
  author={Maury, Bertrand},
  journal={Journal of scientific computing},
  volume={16},
  number={3},
  pages={319--339},
  year={2001},
  publisher={Springer}
}

@article{RBWBM,
title = {Weak impositions of Dirichlet boundary conditions in solid mechanics: A critique of current approaches and extension to partially prescribed boundaries},
journal = {Computer Methods in Applied Mechanics and Engineering},
volume = {348},
pages = {632-659},
year = {2019},
issn = {0045-7825},
author = {Kaizhou Lu and Charles E. Augarde and William M. Coombs and Zhendong Hu},
}

@article{RB1,
title = {Time complexity analysis of quantum difference methods for linear high dimensional and multiscale partial differential equations},
journal = {Journal of Computational Physics},
volume = {471},
pages = {111641},
year = {2022},
issn = {0021-9991},
author = {Shi Jin and Nana Liu and Yue Yu},
}

@article{RB2,
title = {Dense outputs from quantum simulations},
journal = {Journal of Computational Physics},
volume = {514},
pages = {113213},
year = {2024},
issn = {0021-9991},
author = {Jin-Peng Liu and Lin Lin},
}

@article{RB3,
title = {Quantum simulation for partial differential equations with physical boundary or interface conditions},
journal = {Journal of Computational Physics},
volume = {498},
pages = {112707},
year = {2024},
issn = {0021-9991},
author = {Shi Jin and Xiantao Li and Nana Liu and Yue Yu},
}

@article{RB4,
title = {Reduced-order modeling on a near-term quantum computer},
journal = {Journal of Computational Physics},
volume = {510},
pages = {113070},
year = {2024},
issn = {0021-9991},
author = {Katherine Asztalos and René Steijl and Romit Maulik},
}

	\end{document}